\documentclass[draft]{agujournal2019}
\usepackage{url} 
\usepackage{lineno}
\usepackage{ctable}
\usepackage[inline]{trackchanges} 
\usepackage{soul}
\usepackage{graphicx}
\usepackage[utf8]{inputenc}
\usepackage[english]{babel}
\usepackage{multirow}
\graphicspath{ {./} }
\draftfalse

\journalname{Journal of Advances in Modeling Earth Systems}

\begin{document}

%
%

\title{Improving stratocumulus cloud amounts in a 200-m resolution multi-scale modeling framework through tuning of its interior physics}
%
%




 \authors{Liran Peng\affil{1}, Peter N. Blossey\affil{2}, Walter M. Hannah\affil{3}, Christopher S. Bretherton\affil{2,4}, Christopher R. Terai\affil{3}, Andrea M. Jenney\affil{1},Michael Pritchard\affil{1,5}}

\affiliation{1}{Department of Earth System Science, University of California, Irvine, California, USA}
\affiliation{2}{Department of Atmospheric
Sciences, University of Washington, Seattle, Washington, USA}
\affiliation{3}{Lawrence Livermore National Laboratory, Livermore, California, USA}
\affiliation{4}{Allen Institute for Artificial Intelligence, Seattle, Washington, USA}
\affiliation{5}{NVIDIA, Santa Clara, California, USA}





\correspondingauthor{Liran Peng}{liranp@uci.edu}




\begin{keypoints}

\item We improve a long-standing stratocumulus (Sc) dim bias in a high-resolution Multiscale Modeling Framework.
\item Incorporating intra-CRM hypervisocity hedges against the numerics of its momentum solver, reducing entrainment vicinity.
\item Further adding sedimentation boosts Sc brightness close to observed, opening path to more faithful low cloud feedback analysis.


\end{keypoints}

%
%

%
%


\begin{abstract}
High-Resolution Multi-scale Modeling Frameworks (HR) -- global climate models that embed separate, convection-resolving models with high enough resolution to resolve boundary layer eddies -- have exciting potential for investigating low cloud feedback dynamics due to reduced parameterization and ability for multidecadal throughput on modern computing hardware. However low clouds in past HR have suffered a stubborn problem of over-entrainment due to an uncontrolled source of mixing across the marine subtropical inversion manifesting as stratocumulus dim biases in present-day climate, limiting their scientific utility. We report new results showing that this over-entrainment can be partly offset by using hyperviscosity and cloud droplet sedimentation. Hyperviscosity damps small-scale momentum fluctuations associated with the formulation of the momentum solver of the embedded large eddy simulation. By considering the sedimentation process adjacent to default one-moment microphysics in HR, condensed phase particles can be removed from the entrainment zone, which further reduces entrainment efficiency. The result is an HR that can produce more low clouds with a higher liquid water path and a reduced stratocumulus dim bias. Associated improvements in the explicitly simulated sub-cloud eddy spectrum are observed. We report these sensitivities in multi-week tests and then explore their operational potential alongside microphysical retuning in decadal simulations at operational 1.5 degree exterior resolution. The result is a new HR having desired improvements in the baseline present-day low cloud climatology, and a reduced global mean bias and root mean squared error of absorbed shortwave radiation. We suggest it should be promising for examining low cloud feedbacks with minimal approximation.

\end{abstract}

\section*{Plain Language Summary}
Stratocumulus clouds cover a large fraction of the globe but are very challenging to reproduce in computer simulations of Earth's atmosphere because of their unique complexity. Previous studies find the model produces too few Stratocumulus clouds as we increase the model resolution, which, in theory, should improve the simulation of important motions for the clouds. This is because the clouds are exposed to more conditions that make them evaporate away. On Earth, stratocumulus clouds reflect a lot of sunlight. In the computer model of Earth, too much sunlight reaches the surface because of too few stratocumulus clouds, which makes it warmer. This study tests two methods to thicken Stratocumulus clouds in the computer model Earth. The first method smooths out some winds, which helps reduce the exposure of clouds to the conditions that make them evaporate. The second method moves water droplets in the cloud away from the conditions that would otherwise make them evaporate. In long simulations, combining these methods helps the model produce thicker stratocumulus clouds with more water.


%
%

\section{Introduction}

Stratocumulus (Sc) clouds play an important role in the Earth's radiation budget. They are extensive, long-lived, and cool the Earth by reflecting downwelling solar radiation back to space while having little impact on the outgoing longwave radiation. Primarily formed in the presence of large-scale subsidence over cold oceans, the annual mean Sc coverage over the ocean and land is $23\%$ and $12\%$, respectively \cite{wood2012stratocumulus}. Sc also has a strong influence on the heat and moisture exchange between the troposphere and boundary layer \cite{randall1984outlook}. Despite the climatic significance of Sc, climate models do not agree on their seasonal cycle, spatial extent, radiative properties, and cloud feedbacks  \cite{bony2011cfmip,lin2014stratocumulus,gettelman2016processes,brunke2019subtropical,vignesh2020assessment,tselioudis2021evaluation,konsta2022low,zelinka2022evaluating}, and even high resolution models simulate widely varying cloud properties in idealized case studies \cite{ackerman2009large,stevens2005evaluation,bretherton1999intercomparison}. More realistic simulated Sc is necessary to improve our understanding of Sc physics and confidence in projections of the future global-mean temperature \cite{bony2005marine,webb2013origins,dal2015single,tsushima2016robustness,schneider2017earth,zelinka2020causes}. 

Simulating Sc remains a particular challenge because Sc processes involve a wide range of spatial scales and key physical processes that are poorly represented in the subgrid-scale parameterization of global models. Although an Sc cloud deck might cover tens to thousands of kilometers, its thickness is typically only a few hundred meters \cite{wood2012stratocumulus}. Cloud-aerosol interactions are modulated through changes in cloud droplet number, which itself depends on the strength of updrafts whose scale is on the order of 10s or 100s of meters. At the top of subtropical stratocumulus clouds, intense mixing between warm, dry free-tropospheric air and the underlying wet cloud layer occurs within a thin layer typically less than 20 m in vertical extent \cite{ caughey1982field,haman2007small,mellado2017cloud}. \citeA{bretherton2015insights} suggests that such cloud-top entrainment plays a leading role in multiple cloud feedback mechanisms in stratocumulus.  
Because low cloud feedbacks and cloud-aerosol interactions in stratocumulus clouds are thought to be controlled in part by fine-scale processes that are not represented explicitly in storm-resolving models ($\sim$ 1km), this motivates simulations with sub-kilometer grid spacing \cite<e.g., >{stevens2020added}.     

Several interesting strategies have emerged in recent years to capture more explicit and plausible stratocumulus dynamics in next generation global climate models. First, \citeA{lee2022resolving} demonstrate some potential from adaptively refining vertical grid structure solely within a strategic subset of the physical parameterization suite, with higher order closure scheme used in Cloud Layers Unified By-Binormals (CLUBB, \citeA{golaz2002pdf2,golaz2002pdf,larson2005using,larson2012pdf}) and vertical transport (e.g. subsidence and sedimentation/precipitation). This has advantages of producing some better baseline Sc \cite{bogenschutz2013higher,lee2022resolving}, but disadvantages of accepting all the limitations of operational subgridscale turbulence schemes. Second, \cite{lopez2020generalized} sidesteps the turbulence parameterization problem by using very highly resolved ($\Delta x$ = $\Delta y$ = 35 m and $\Delta z$ = 5 m) three dimensional LES, managing computational expense by using a sparse ensemble as a library from which to train eddy diffusivity / mass flux based parameterization schemes \cite{cohen2020unified}. Advantages of the highly resolved LES choice include a luxuriously converged limit that sidesteps most need to parameterize beyond microphysics; disadvantages include imposing idealizations of lateral periodicity and a scale separation in their harness to a global host, as well as limited geographic sampling due to the expense of such LES; however, the latter is positioned to be managed with  calibration schemes that may inform where such calculations can be strategically deployed to maximum global benefit \cite{dunbar2022ensemble}. Advantages of the Eddy-Diffusivity Mass-Flux (EDMF) framework include its interpretability and generalizability; disadvantages include its potential inability to subsume some complicated organization feedbacks. Finally, \citeA{miyamoto2013deep} avoids scale separations entirely by directly resolving fully global uniform 870-m horizontal resolution. Similar work currently planned to attain 200-m global horizontal resolution has advantages of resolving the outer scale of boundary layer eddies without drawing scale separations, but the disadvantage of the inordinate computational expense and inability to conduct multidecadal cloud feedback experiments, as well as a limited ability to refine vertical grids near the inversion. All of the above approaches must cope with the ongoing difficulties and uncertainty of how to parameterize microphysics.

We will focus on a strategy that is complementary to all the above approaches for dealing with the computational challenge of low clouds for climate simulation, by using the multiscale modeling framework (MMF, also referred to as ``superparameterization (SP)''; \citeA{grabowski2004improved,khairoutdinov2001cloud,hannah2020initial}), in which a coarse resolution ($\sim$100~km) global climate model (GCM) is coupled to an embedded convection-resolving models (CRMs) at each global grid location. Many previous studies ~\cite{kooperman2016impacts,kooperman2016robust} have shown low-resolution MMF (LR) tests with a traditional MMF, i.e. using coarse 4-km horizontal resolution and greater than 100-m vertical spacing, can improve the simulated rainfall distribution and wave spectrum near the equator and over summer continents. However low-cloud-forming eddies are not directly resolved in this approach, requiring subgrid scale parameterizations to cope with \cite{wang2015multiscale,cheng2015improved}. 

The MMF coupling paradigm does not put constraints on the CRM domain size or grid spacing, so large eddy permitting resolution can be used to sidestep most of the parameterization problem regarding cloud-forming eddies. The MMF approach makes the physical idealizations of imposing lateral periodicity in the CRM and a scale separation between the GCM and CRM, and historically has limited the CRM to just two dimensions for computational efficiency. Despite these concessions the MMF has a unique computational advantage that enables full geographic sampling and relatively fast throughput, with options to accelerate the CRM algorithmically \cite{Jones2015mean-state-accel} or via GPU hardware \cite{hannah2020initial}, and regionalized load balancing \cite{peng2022load} that makes the MMF approach increasingly attractive for climate dynamics and low cloud feedback applications. 

A high-resolution MMF (HR) with a grid designed for low cloud simulations was first explored by ~\citeA{parishani2017toward} with hopes of more directly simulating shallow convection over Sc-covered regions. Encouraging initial improvements in low cloud vertical structure, diurnal sensitivity, and the vertical structure of sub-cloud turbulent kinetic energy were reported in a model configuration using simplified bulk, one-moment cloud microphysics.

However, such HR experiments have to date been unable to sustain sufficient liquid water in stratocumulus regions, where simulations suffer from an undesired regional dim bias that has been difficult to overcome ~\cite{parishani2017toward}. Associated symptoms have implicated an unknown source of vertical mixing that disrupts the balance required to sustain morning Sc by mixing too much free tropospheric air into the boundary layer. The overall impact is to under-predict daytime cloud liquid water resulting in too little time mean shortwave reflectivity. Meanwhile, the assumptions inherent in an MMF that can limit its ability to laterally advect condensed water between adjacent CRMs have caused some to question its capacity to maintain low clouds \cite{jansson2022representing}. While the scale separation inherent in the MMF also introduces distortions (such as the neglect of the mesoscale), it does allow a global model to simulate these fine scales, making it possible to represent physical processes (e.g., cloud top entrainment, aerosol activation in updrafts) that drive critical sensitivities of low clouds to anthropogenic influence.

In short, the question is open as to whether the HR approach should ever be expected to maintain realistic amounts of liquid in marine Sc regions, to the extent that it must rely primarily on local cloud generation to succeed, and given that over-entrainment has proved a stubborn problem to overcome. Motivated by the Transpose-Atmospheric Model Intercomparison Project \cite<Transpose-AMIP, >{williams2013transpose}, we use the hindcast approach to test different model configurations. In this context, the purpose of this paper is to explore two mechanisms to control entrainment efficiency in a HR and examine the extent to which they can alleviate the Sc dim bias issue. The first is to numerically damp unphysical noise at the grid scale caused by the numerics by applying a hyperdiffusive term (which we will refer to as ``hyperviscosity'') to the momentum equation that can reduce entrainment and entrainment efficiency \cite{wyant2018sensitivity}. Second, enhancing cloud droplet sedimentation (henceforth, ``sedimentation''), which can also reduce the entrainment efficiency and preserve cloud liquid by depleting the liquid water in the cloud-top entrainment zone \cite{bretherton2007cloud}. We will use the term "entrainment efficiency" in our study as an indicator of the magnitude of tendencies resulting from the combined effects of explicitly resolved mixing, numerical diffusion, and parameterized subgrid-scalediffusion.

The paper is organized as follows: Section 2 contains a rationale and description of how we implement both hyperviscosity and sedimentation processes. In Section 3, we first analyze the results of six, short-duration sensitivity simulations in a testbed HR configuration to understand the impact of varying degrees of CRM-scale hyperviscosity and sedimentation on low cloud characteristics and the spectrum of turbulent eddies in the marine boundary layer. These results point to temporal nonlinearities and a promising configuration for longer-duration simulations in an operational configuration, for which we show results from a subsequent round of microphysical tunings. This allows an updated view of HR top of atmosphere radiative biases after controlling for over-entrainment. A summary of the results and a discussion are included in section 4. 

 \section{Methods}\label{Method}

\subsection{Model Description}\label{ModelDes}

In this study, we use the Multi-scale Modeling Framework configuration of the Energy Exascale Earth System Model (E3SM-MMF; ~\citeA{hannah2020initial}) as a testbed to examine the impact of hyperviscosity and sedimentation on low clouds simulated by high resolution embedded convection arrays. E3SM was forked from the NCAR CESM \cite{Hurrell2013TheResearch} but has undergone continued development and enhancement since then \cite{golaz2019doe, Xie2018UnderstandingModel}. The dynamical core uses a spectral element method on a cubed-sphere geometry \cite{ronchi1996cubed,taylor2007mass}. Physics calculations are done on a finite volume grid that is slightly coarser than the spectral element grid used for dynamics, but the physics grid is comparable to the effective resolution of the dynamics grid and does not alter the qualitative behavior of the model \cite{hannah2021separating}.

Each simulation follows the approach of \cite{khairoutdinov2005simulations} with a two-dimensional CRM based on the System for Atmospheric Modeling \cite<SAM, >{khairoutdinov2005simulations} embedded within each GCM physics column. These embedded CRMs are oriented meridionally within the host GCM grid cell and have periodic lateral boundary conditions. The vertical grid and background anelastic state are updated to match the parent GCM column for each CRM integration (typically once per GCM time step). The CRM uses a one-moment microphysics scheme with a temperature-dependent partitioning of the cloud condensate (cloud water and ice) and a subgrid-scale (SGS) turbulent transport scheme with a diagnostic Smagorinsky closure. The rapid radiative transfer model for General model application (RRTMGP) \cite{pincus2019balancing} is used for radiation, which has been rewritten in C++ to run efficiently on GPUs. CRM columns are combined for radiative calculations to reduce the computational burden of radiation instead of considering each CRM column separately, which does not qualitatively affect the model solution for typical configurations with 4 or more radiative columns. Radiative tendencies are calculated once each GCM time step and are applied back to the corresponding group of CRM columns on the following time step. The domain average CRM variables for temperature and water species are used to calculate forcing and feeedback tendencies in order to couple the CRM and GCM, following the conventional MMF coupling scheme described by \citeA{grabowski2004improved}. 

 \subsection{Hyperviscosity}\label{Hyperviscosity}

We now proceed to outline the first method envisioned to control MMF stratocumulus entrainment in HR configurations. Models like SAM use oscillatory centered difference numerical schemes \cite{wicker2002time} for momentum advection and upstream biased discretizations of scalars with SGS closures have relatively weaker performance than other numerical formulations in an LES case study based on the first research flight (RF01) of the second Dynamics and Chemistry of Marine Stratocumulus (DYCOMS-II) field campaign \cite{pressel2017numerics}. \citeA{wyant2018sensitivity} find that in 3D LES simulations with a horizontal resolution of 35 m and vertical grid spacing as fine as 5 m, a hyperdiffusion term helps to increase LWP by numerically damping small-scale eddies and reducing entrainment and entrainment efficiency. We test this method in HR's 2D CRM arrays by applying a hyperdiffusive term \cite{wyant2018sensitivity} to the momentum equation inside the CRM. 

The fourth-order hyperdiffusivity can be written as 
\begin{equation} \label{eq:Hyper}
\partial _{t} {\vec{u}} =  -k \nabla_h^{4} \vec{u},
\end{equation}
\noindent where $k=\Delta x^4/(16 \, \tau)$ is the effective diffusivity and $\nabla_h^{4}$ operator applies along the horizontal ($x$) direction of our 2D CRM using the fourth derivative central finite difference with second-order accuracy. Compared to Laplacian diffusivity $\nabla^{2}$, hyperdiffusivity can more selectively damp the smallest-scale structures confined in smaller wavelength ranges, with little impact on larger scales \cite{maron2008constrained}. This value of $k$ damps oscillations with Nyquist wavelength on a time scale of $\tau$, which has a default value of 30~s in our simulations. 

It is important to note that in LR configurations with $\Delta x = 1200$ m and a Nyquist wavelength of 2400 m, such hyperviscosity should be expected to be counterproductive, given that LR low clouds rely on under-resolved grid-scale ``eddies'' to deliver moisture from the surface; in this context, hyperviscosity should be expected to shut down low cloud formation unhelpfully, something we have confirmed (not shown). But in our HR configurations with $\Delta x = 200$ m the use of a filter like Equation \ref{eq:Hyper} is better posed given that the cloud-forming boundary layer eddies occupy multiple horizontal grid columns. Put another way, only as MMFs have exited their infancy to allow sub-km horizontal resolution, has CRM-scale hyperviscosity become an interesting consideration. 

\subsection{Sedimentation}\label{Sedimentation_Des}

Our second method to control entrainment efficiency considers a slight modification of HR's simple microphysics. In the one-moment microphysics scheme used in our simulations, condensation occurs when the water vapor amount exceeds saturation, with the excess above saturation converted to liquid, ice, or a mixture of the two depending on temperature. While precipitating liquid and ice (rain, snow, and graupel) sediment as described in \citeA{khairoutdinov2003cloud} and \citeA{heymsfield2003properties}, cloud liquid droplets did not sediment in this scheme.  Such droplets do sediment in reality and the inclusion of sedimentation in numerical models leads to liquid water retention and increased in-cloud liquid water content \cite{bretherton2007cloud}. Here, we follow  \citeA{yau1996short} and define the precipitation flux of cloud liquid due to sedimentation as
\begin{equation} \label{eq:Sedi}
P = c[3/(4\pi \rho_{l} N_{d})]^{2/3}(\rho q_{c})^{5/3}exp(5ln^{2} \sigma_g),
\end{equation}
\noindent where $\rho$ is the air density of air, $\rho_{l}$ the water density, $N_d$ the cloud droplet number concentration, $q_{c}$ the cloud liquid water mixing ratio, $\sigma_g$ the geometric standard deviation of the (lognormal) cloud droplet size distribution, and $c=1.19\times 10^8$ $m^{-1}s^{-1}$. The cloud droplet number concentration is prescribed as a constant 70 $cm^{-3}$ over ocean and 140 $cm^{-3}$ over land. A larger value of $\sigma_g$ corresponds to a broader size distribution and a faster terminal velocity for larger size droplets.  \citeA{geoffroy2010parametric} provide estimates of $\sigma_g$ based on two marine field campaigns, with a central estimate of 1.34 and a parameterization of $sigma_g$ with values ranging from 1.2 to 1.5 for liquid water contents ranging from 0.01 to 2 g m$^{-3}$.  Below, experiments with fixed values of $\sigma_g =$ 1.2 and 1.5 are used to characterize the impact of sedimentation in HR MMF.

The cloud optical depth is closely related to the cloud fraction, effective cloud droplet radius, and the in-cloud liquid/ice water content. Following the previous implementation of the single-moment microphysics, fixed values of cloud effective radius for land and ocean are used for all simulations. Only cloud fraction and liquid/ice water content  affect cloud optical properties. As we will see, incorporating these effects of sedimentation will reduce entrainment efficiency by drawing liquid down from the inversion zone, especially for larger values of $\sigma_g$.

 \subsection{Experimental Design}\label{Experiments}

By default, E3SM-MMF uses a $60$ level vertical grid (L60), which is coarser than the default $72$ level grid (L72) used by E3SM. The L72 grid was implicated as the cause of intermittent numerical instability in E3SM-MMF due to very thin layers near the surface, which is often around $20$ m thick in the lowest layer. Instead of addressing this by reducing the time step the L60 grid was designed to avoid instability with an approximate thickness of $100$ m in the lowest level. Alternatively, the simulations presented here utilize a $125$ level grid (L125) that is designed to concentrate refinement roughly between $500$ and $1800$ m to improve the representation of sharp temperature inversions needed to represent marine stratocumlus clouds. A smaller CRM time step is used to avoid any numerical issues. Thus, the configuration referred to as ``LR'' in this study is not the classical cloud SP that uses both coarse vertical and horizontal resolution; rather it can be compared to the ``C32-L125-250m'' MMF grid configuration of \citeA{parishani2017toward}. According to \cite{bretherton1999intercomparison}, this vertical grid spacing is not sufficient to resolve entrainment but reflects a pragmatic choice that is computationally affordable in the E3SM-MMF and is intentionally consistent with \citeA{parishani2017toward}.

In Sections \ref{Results_Ghindcast} and \ref{Turbulence_scale}, a computationally efficient configuration will be exploited for hindcast experiments by using a relatively coarse ne16pg2 global grid with 6,144 columns for physics calculations (approximately $2.8$ degree grid spacing). These simulations use a $10$ min GCM physics time step. To ensure computational efficiency, the radiation calculations were constrained to $16$ columns. This means that radiative heating was computed based on time and spatial averages, considering evenly grouped CRM columns within each GCM column. Each simulation was run for $15$ days starting from an initial condition derived from European Centre for Medium-Range Weather Forecasts (ECMWF) Reanalysis (ERA5) atmospheric data \cite{hersbach2020era5} and National Oceanic and Atmospheric Administration (NOAA) sea surface temperature and sea ice data. All hindcast simulations are initialized from 1 October 2008. In our LR hindcast configuration, the embedded CRM has $32$ columns with a horizontal grid spacing of $1200$ m ($38.4$ km extent) and a $5$ s CRM time step (see Table \ref{tab:exp_summary}). Our HR hindcast configuration we use $64$ columns with a horizontal grid spacing of $200$ m ($12.8$ km extent) and a $0.5$ s CRM time step. 

\begin{table}
 \caption{A summary of the simulations performed in this study}
 \centering
 \begin{tabular}{c c c c c c c c c}
 \hline
   Simulation ID & levels & N & $dx$ (m) & Extent (km) & $dt$ (s)& $\tau$ (s) & $\sigma_g$\\
 \hline
   LR                        & 125 & 32 &  1200 & 38.4 &   5&  - &   -\\
   HR                        & 125 & 64 &   200 & 12.8 & 0.5&  - &   -\\
   HRh                      & 125 & 64 &   200 & 12.8 & 0.5& 30 &   -\\
   HRh15                    & 125 & 64 &   200 & 12.8 & 0.5& 15 &   -\\
   HRs15                & 125 & 64 &   200 & 12.8 & 0.5& - & 1.5\\   
   HRhs12                & 125 & 64 &   200 & 12.8 & 0.5& 30 & 1.2\\
   HRhs15                & 125 & 64 &   200 & 12.8 & 0.5& 30 & 1.5\\
 \hline
 \label{tab:exp_summary}
 \end{tabular}
 \begin{tablenotes}
      \small
      $\bullet$ Note: $N$=number of CRM columns; $dx$=CRM horizontal resolution; $\tau$=damping time scale; $dt$= CRM time step; $\sigma_g$=the logarithmic width of the droplet size distribution, for the simulations that included sedimentation effects.
 \end{tablenotes}
 \end{table}
 
The first two hindcast experiments are used to compare the LR and HR configurations (first two rows of Table \ref{tab:exp_summary}). The rest of the hindcasts are based on the HR configuration and perturb the magnitude of $\tau$ in Eq.\ref{eq:Hyper} and $\sigma_g$ in Eq.\ref{eq:Sedi}. In HRh, we add hyperdiffusion, with $\tau = 30$ s. In HRh15, we test the model sensitivity to the damping time scale with $\tau = 15$ s. Halving $\tau$ doubles the magnitude of $k$ in Eq.\ref{eq:Hyper}, which intensifies the damping of small-scale turbulent eddies; this will turn out to have some encouraging but insufficient improvements in low cloud amount. The HRhs12 and HRhs15 simulations combine hyperdiffusion (with $\tau$ fixed at 30 s) with perturbed cloud drop size distributions using $\sigma_g = 1.2$ and 1.5, respectively. As we will see, it is these latter experiments that produce the most encouraging improvement in the stratocumulus dim biases that have hampered past incarnations of HRh. 

In section \ref{MicroTuning}, we explore more computationally abitious simulations using a ne30pg2 global grid with 21,600 physics columns (approximately $1.5$ degree grid spacing). These simulations are similar to the HR configuration described above but with several notable differences in their configuration, specifically a $20$ min GCM physics time step, a $2$ s CRM time step, $256$ CRM columns with a horizontal grid spacing of $200$ m ($51.2$ km domain extent), and $4$ radiative columns. Another important difference of these runs is that they utilize schemes for convective momentum transport \cite{tulich2015strategy, yang2022E3SM-MMF-CMT} and CRM variance transport \cite{hannah2022-variance-transport}, which have recently been shown to improve various aspects of E3SM-MMF. Each tuning experiment was run for six months, from January to June, using seasonally-varying climatological conditions based on the years 2005--2014. 
This ambitious ensemble was made possible by ongoing development to enhance the throughput of E3SM-MMF, which includes code refactoring to leverage GPU hardware acceleration on the Oak Ridge Leadership Computing Facility (OLCF) Summit machine \cite{Norman2019UnprecedentedSupercomputer}. 

\begin{figure} 
    \centering
    \includegraphics[scale=0.5]{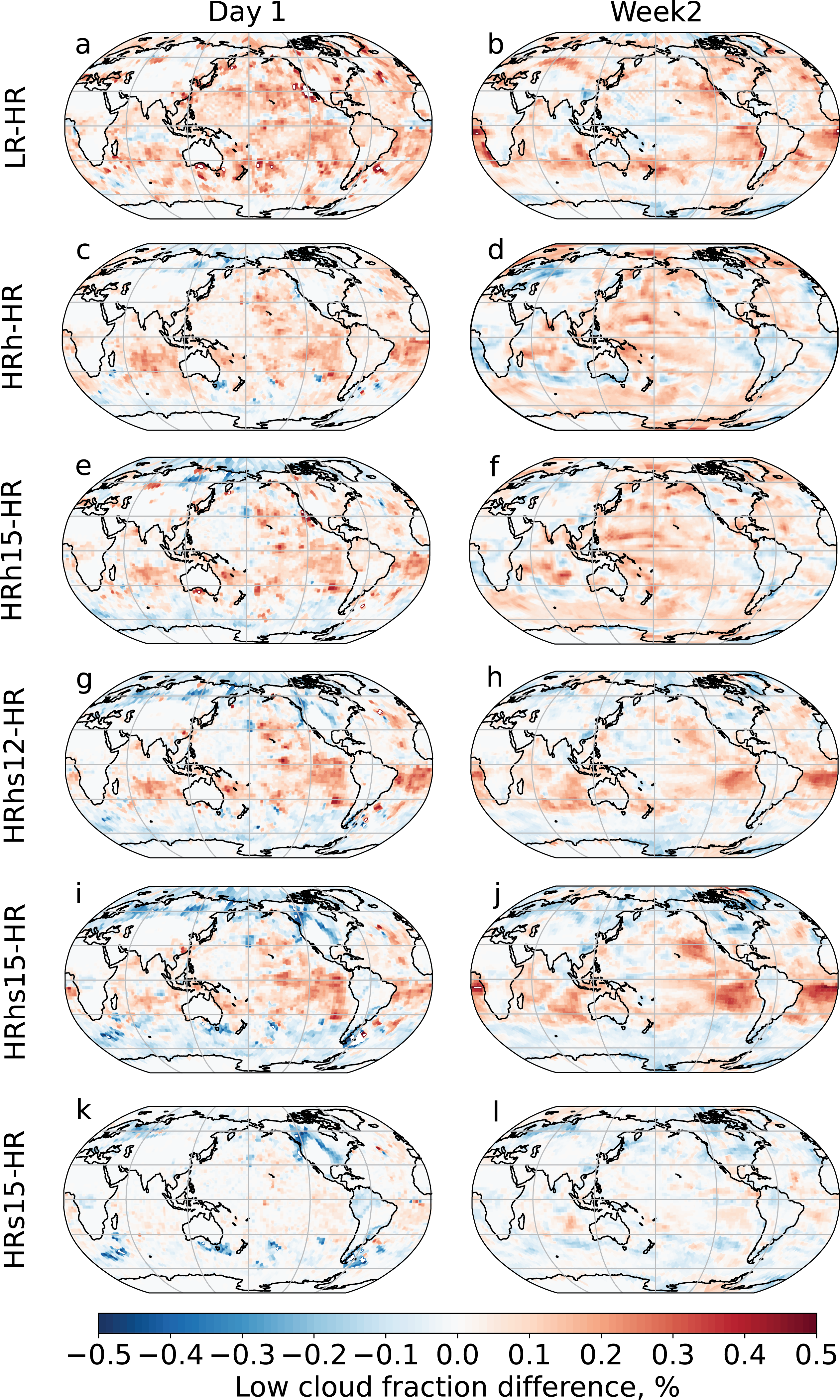}
    \centering
    \caption{Low cloud fraction differences based on the first day average (left) and two-week averaged (right) between (ab) LR and HR (cd) HRh and HR, (ef) HRh15 and HR, (gh) HRhs12 and HR, (ij) HRhs15 and HR, and (kl) HRs15 and HR}
    \label{fig:lcc}
\end{figure}

 \section{Results}\label{Results}

 \subsection{General Features of the Simulation Based on Global Hindcast Results}\label{Results_Ghindcast}

Because the HR configuration has been identified as suffering from a deficit of low cloud in stratocumulus regions (by as much as 20\% in the Sc covered ocean), we simulate two week hindcasts from October 2008 with the model configurations in Table~\ref{tab:exp_summary} and seek those configurations that produce sustained more low cloud relative to HR.  Figure \ref{fig:lcc} shows the change in low cloud fraction relative to HR for each model configuration, with results from both the first day and the second week of the simulations used to identify the initial and longer-term responses. As in \citeA{parishani2017toward}, the LR configuration produces increased cloud cover relative to HR, with widespread increases that are not focused in the stratocumulus regions (Figure \ref{fig:lcc}a,b). The first encouraging result is that selectively damping small size eddies can retain more subtropical stratucumulus clouds during the first simulated day than HR (geographic patterns in Figure \ref{fig:lcc}c,e), but unfortunately that improvement is transient so that this initial effect is not sustained over a two-week average (Figure \ref{fig:lcc}d,f). While enabling sedimentation of cloud droplets provides modest ($\sim$0.1\%) increases in low cloud in some stratocumulus regions (Figure \ref{fig:lcc}k,l), \textit{combining} sedimentation with hyperviscosity leads to an even stronger initial stratocumulus cloud increase (Figure \ref{fig:lcc}g,i) that is sustained on longer timescales (Figure \ref{fig:lcc}h,j), suggesting a viable path to improve HR's climatological stratocumulus bias. Increasing the droplet size broadness parameter amplifies this effect (Figure \ref{fig:lcc}j). 

If one estimated the low cloud change with the combined effects of hyperviscosity and sedimentation in HRhs15 (Figure ~\ref{fig:lcc}j) as the sum of their individual impacts in HRh and HRs15 (Figure ~\ref{fig:lcc}d,l), the estimate differs greatly from the result in HRhs15 in both its magnitude and in the regional distribution of cloud changes.  This nonlinear response in HRhs15 concentrates low cloud increases in the stratocumulus regions, suggesting the synergistic interactions of hyperviscosity and cloud droplet sedimentation lead to more persistent stratocumulus clouds.  We believe that these clouds are sustained by more realistic turbulent circulations within the marine boundary layer and will explore this further in section~\ref{Turbulence_scale}.

\begin{figure} 
    \centering
    \includegraphics[scale=0.5]{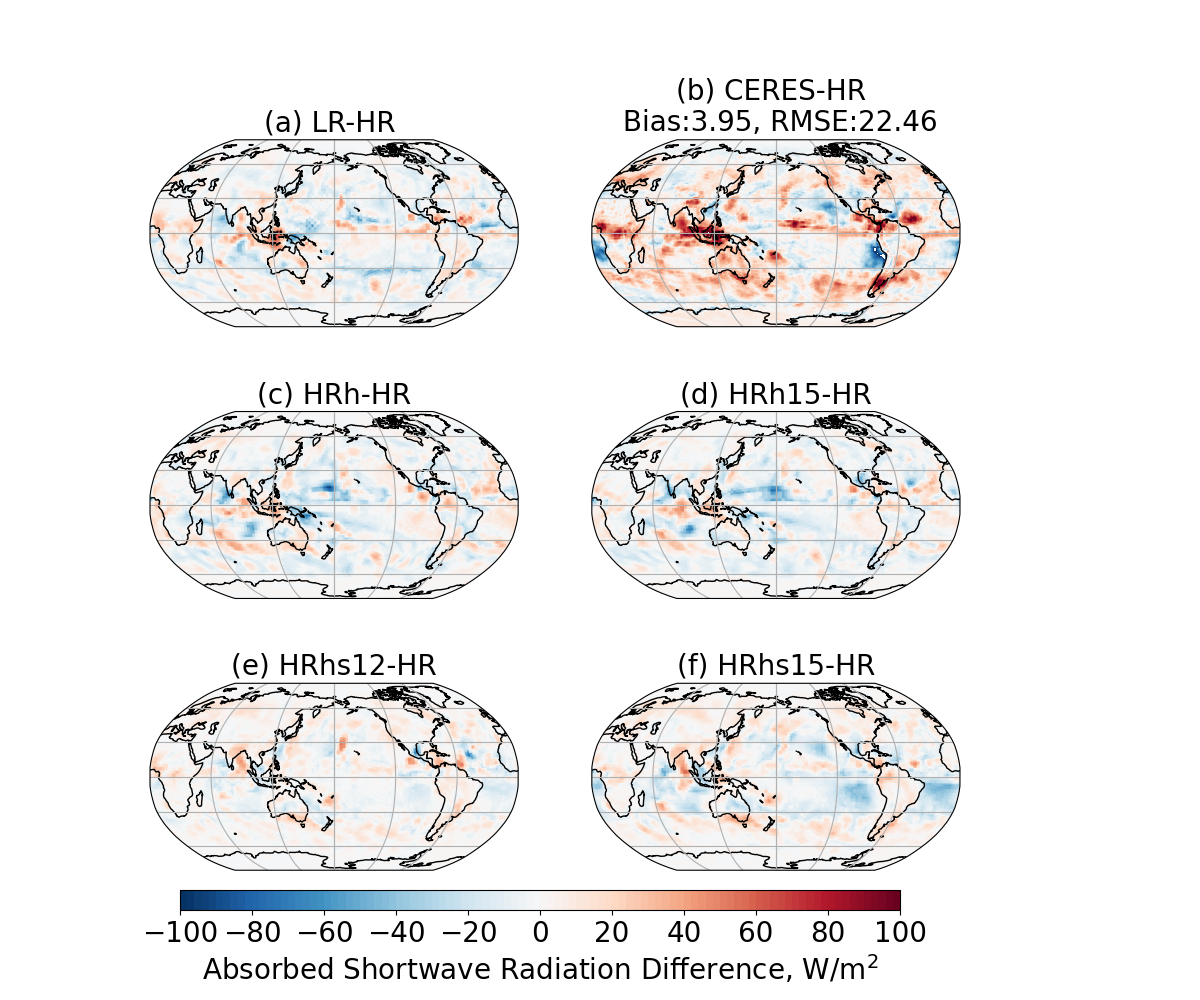}
    \centering
    \caption{Averaged absorbed shortwave radiation (ASR) differences between (a) LR and HR, (b) CERES and HR, (c) HRh and HR, (d) HRh15 and HR, (e) HRhs12 and HR, and (f) HRhs15 and HR. }
    \label{fig:asr}
\end{figure}

We next examine whether these cloud fraction increases are radiatively significant relative to the shortwave stratocumulus dim bias of concern. Based on the two-week averages, Figure \ref{fig:asr}b first calibrates its structure and magnitude relative to CERES data in our simulations: Note the collection of negative ASR anomalies over the Sc regions makes a large contribution to the RMSE. Unlike other regional details of the absorbed shortwave radiation (ASR) biases that are difficult to disentangle from internal variability in a 15-day sample, the consistent ASR bias over both of the subregions of most subtropical Sc during October (off the western coasts of Namibia and Peru) indicates a robust climatological signal. Despite using a different dynamical core and modeling framework from \citeA{parishani2017toward}, this baseline ASR bias is similar to their C32-L125-250m simulation (their Figure 4b) indicating some stable signals across different MMF implementations. 

We now look at change in ASR from this baseline for the simulations with hyperviscosity and sedimentation (Figure \ref{fig:asr}c-f). Consistent with its effects on cloud fraction, when hyperviscosity is used in isolation (both HRh and HRh15; Figure \ref{fig:asr}c-d) there is no reduction in the dim bias over Sc regions; rather, the dim bias becomes slightly more pronounced in those simulations. When both effects are combined, cloud induced brightening occurs throughout the stratocumulus regions (relative to HR), except in the near-coastal environment, acting encouragingly in the same direction as the baseline anomalies (negative ASR anomalies off the west coasts of the Namibia, Peru, and Western Australia in Figure \ref{fig:asr}f). This reduced ASR dim bias corresponds well with the locations of low cloud fraction increase (Figure \ref{fig:lcc}hj). None of the four model perturbations introduce notable outgoing longwave radiation (OLR) differences compared to HR based on the 15-day means (Figure~\ref{fig:olr}). 

 \begin{figure} 
    \centering
    \includegraphics[scale=0.5]{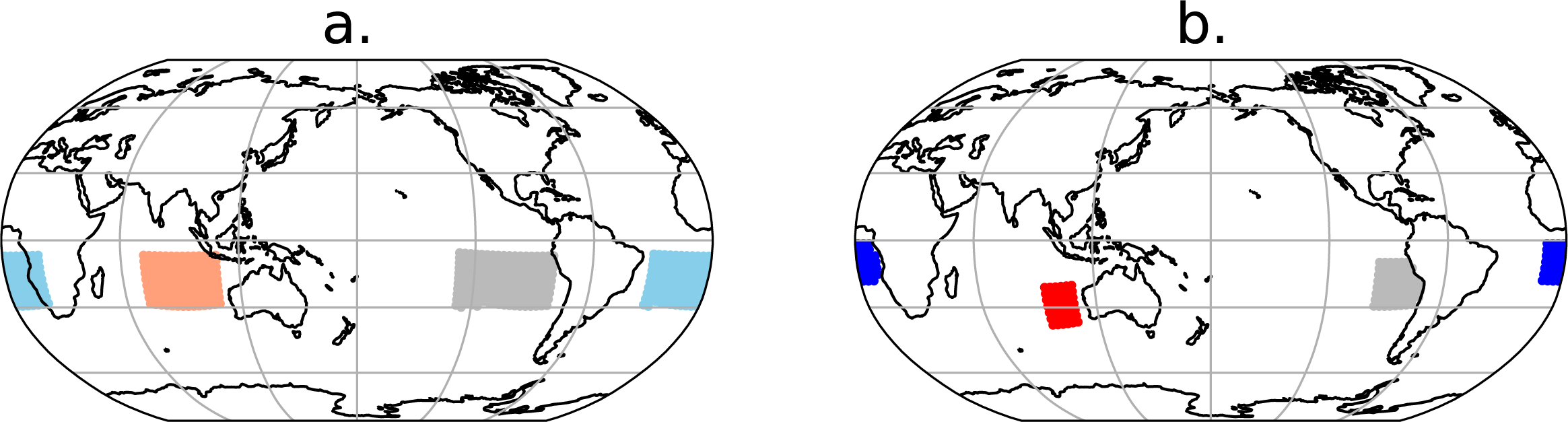}
    \centering
    \caption{The map of (a) highlighted regions used in 
    Figure  \ref{fig:asrdiff} and (b) the selected regions to construct height-time plots over Peruvian (gray), West Australian (red), and Namibian (blue). }
    \label{fig:Peruvianmask}
\end{figure}

 \begin{figure} 
    \centering
    \includegraphics[scale=0.25]{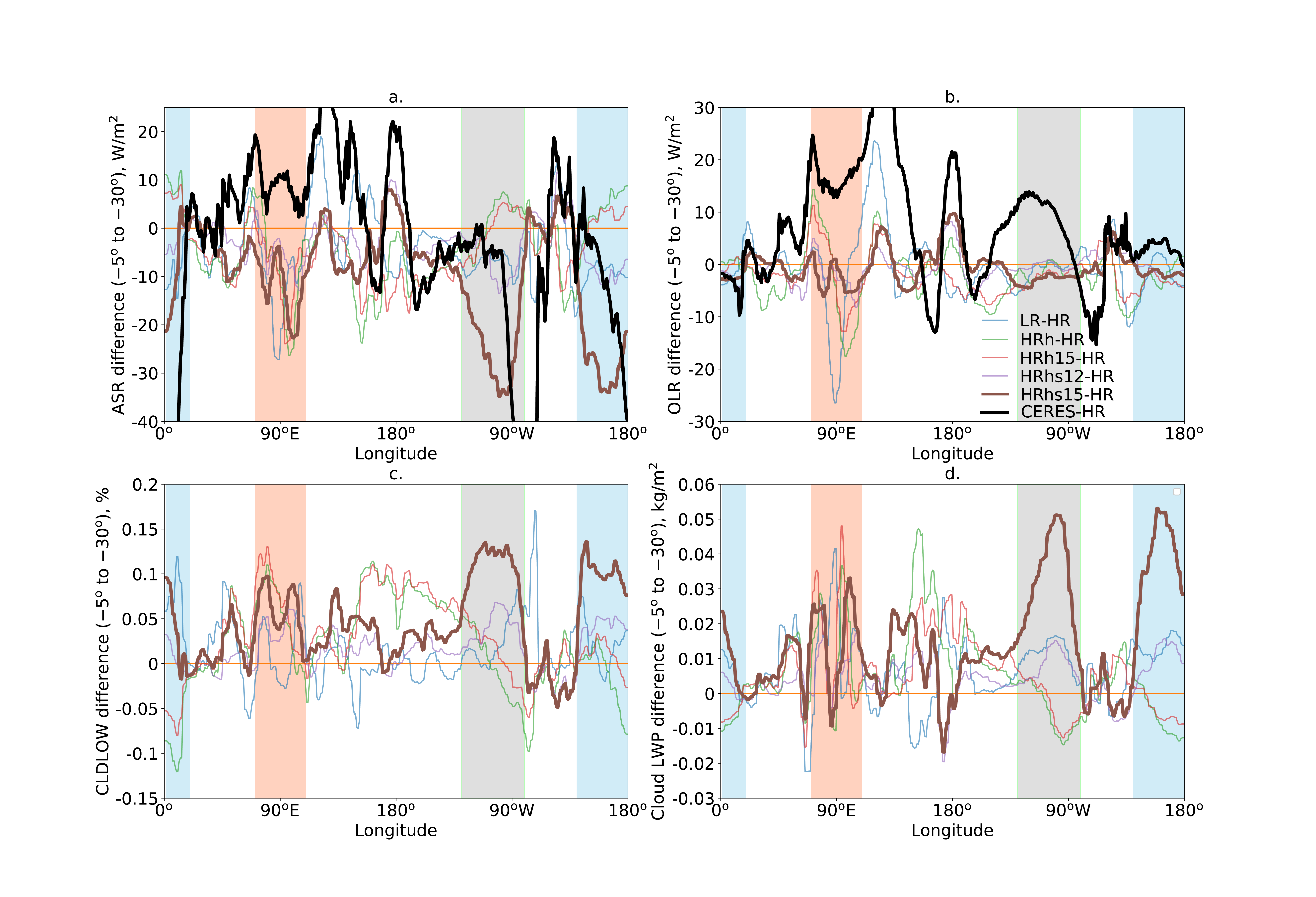}
    \centering
    \caption{A comparison of (a) the meridional mean absorbed shortwave radiation (ASR) (b) outgoing longwave radiation (OLR), (c) low cloud fraction, and (d) cloud liquid water path (LWP) differences between LR and HR, HRh and HR, HR15 and HR, HRhs12 and HR, HRhs15 and HR, and CERES and HR. The three Sc regions, namely the west coast of Australia, Peru, and Namibia, are represented by the orange, gray, and blue shadings from left to right, respectively.}
    \label{fig:asrdiff}
\end{figure}

We now hone in on a strategic subregion of the Southern Hemisphere subtropical Sc latitudes (highlighted in Figure \ref{fig:Peruvianmask}a). Inter-model differences of meridionally and time average properties along this zonal transect are shown in Figure \ref{fig:asrdiff}. The strongest and most interesting changes relative to HR occur when hyperviscosity and sedimentation are combined in HRhs15, producing encouraging Sc brightening that is emphasized in the thick, dark red line. The thick black line serves as a reference, representing the comparison between CERES data and the baseline HR simulation. Closeness to this line - as begins to occur in the eastern parts of the Peruvian (grey) and Namibian (blue) regions, indicates a reduced bias. To orient, the shaded regions (Figure \ref{fig:asrdiff}) delineate the three Sc zonal subregions highlighted in Figure \ref{fig:Peruvianmask}a: off the west coasts of Australia (orange shading), Peru (gray shading), and Namibia (blue shading); in these regions, HRhs15 produced time-mean cloud brightening relative to HR on the order of 20-35 W/m$^2$. The other panels show that in these same regions HRhs15 also produce 5-13 $\%$ more low cloud (Figure \ref{fig:asrdiff}c) and 0.03-0.05 $kg/m^{2}$ larger cloud liquid water paths (Figure \ref{fig:asrdiff}d). When hyperviscosity is used in isolation (HRh and HRh15), cloud brightening occurs over the Australian Sc deck relative to HR but the dim bias is worsened over the Peruvian and Namibian Sc decks; that is, the effect is not systematic across Sc regimes. Likewise, HRh and HRh15 have either roughly no change or reduced cloud amount relative to HR and liquid water path for the two regions where the dim bias worsens.  

In summary, we find robust reductions in the two-week-mean HR dim bias over the Sc regions in HRhs15 which occur due to increased cloud amount and liquid water path. Changes seem to be due mainly to the synergistic effects of sedimentation and hyperviscosity, as their combined effect is much larger than when either is used in isolation.

\begin{figure} 
    \centering
    \includegraphics[scale=0.3]{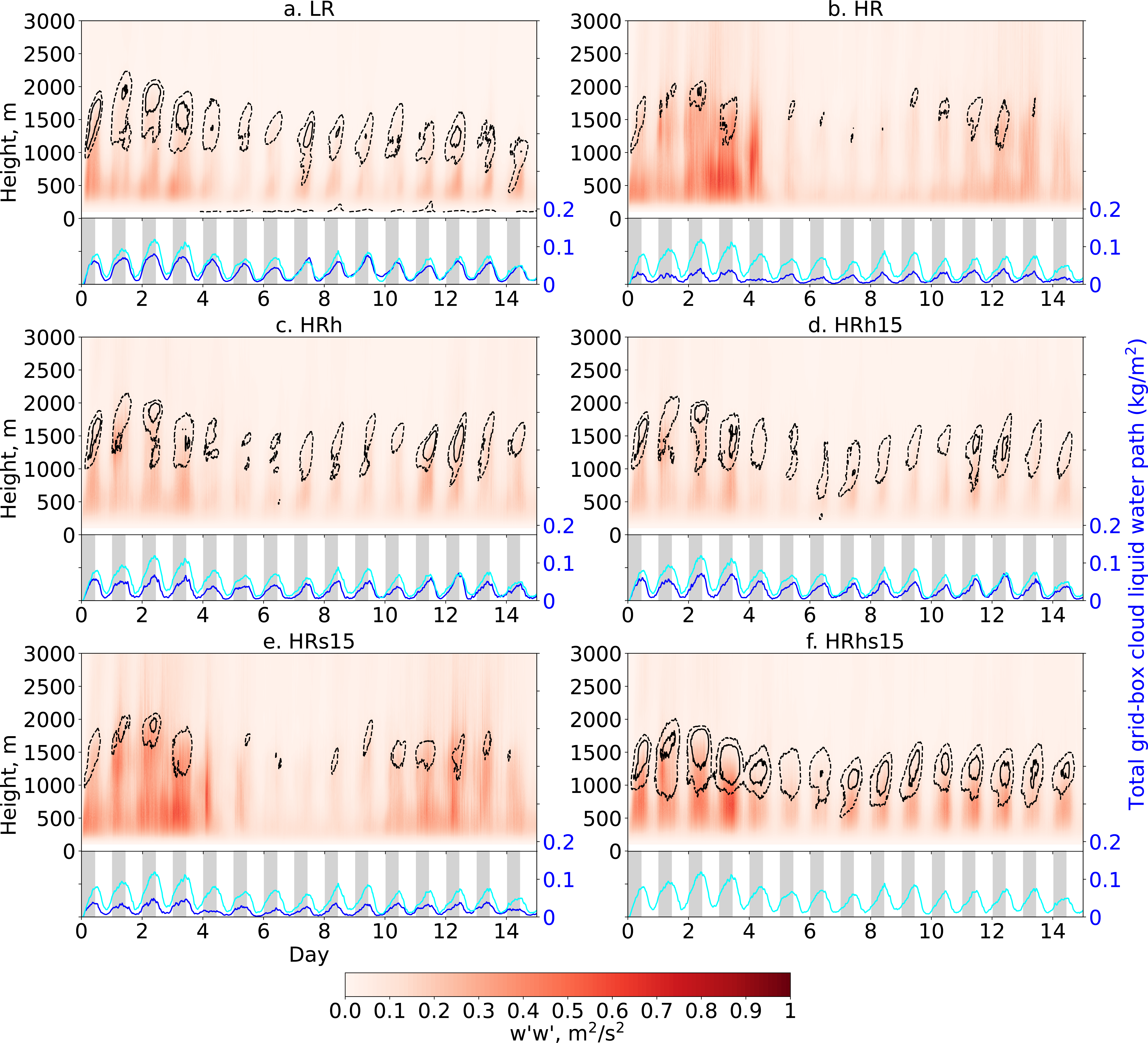}
    \centering
    \caption{Height time evolution of vertical velocity variance ($\overline{w'w'}$, in units of $m^2/s^2$) in Peruvian averaged over 15 days starting from October 1st 2008. (a) LR, (b) HR, (c) HRh, (d) HRh15, (d) HRhs12, (f) HRhs15. The blue lines represents the total grid-box liquid water path and cyan line represents HRhs15 for all panels as a reference. Gray shaded time intervals represent nighttime. Black contours are showing the $10\%$ (black dashed line) and $20\%$ (black solid line) cloud fraction. }
    \label{fig:htww}
\end{figure}

 \subsection{Analysis of Peruvian Stratocumulus Region}\label{Stratocumulus_Region}

This section focuses on the Peruvian Sc region to examine further details of the unsteady evolution of boundary layer vertical structure and the associated changes in low clouds. This region, lying off the west coast of South America over the ocean (gray area in Figure \ref{fig:Peruvianmask}b), is one of the most persistent Sc decks \cite{bretherton1997moisture} and poorly simulated in models \cite{konsta2022low}. 

Figure \ref{fig:htww} shows the time-height evolution of CRM-scale vertical velocity variance (shading), a good proxy of low-level turbulent mixing, revealing its co-evolution with cloud fraction (black contours). Both quantities are averaged over the Peruvian Sc region (Figure ~\ref{fig:Peruvianmask}b), as is the time series of liquid water path shown below each contour plot.  The cyan line benchmarks the liquid water path in HRhs15, the simulation that resulted in the most liquid. Strong diurnal cycles are apparent in all simulations, with strong turbulent mixing occurring during local nighttime, as expected \cite{hignett1991observations}. Compared to LR, which uses embedded CRMs that are larger and have a much coarser horizontal resolution, the baseline HR (Figure \ref{fig:htww}b) produces a larger magnitude of $\overline{w'w'}$ which also extends to a higher altitude, at least during the first few simulated days. These signals and differences between HR and LR are consistent with \citeA{parishani2017toward}, including the inability of HR to sustain low clouds beyond day 4, consistent with its dim bias. Interestingly, in HR some nontrivial $\overline{w'w'}$ is found above the cloud layer, whereas in LR, above-cloud $\overline{w'w'}$ is near-zero. The $\overline{w'w'}$ vertical structure for HR simulation appears to have two modes during the first two days of simulation, with one mode near the surface and the other mode closer to the cloud layer, suggesting decoupling. On the other hand, LR only shows one local maximum $\overline{w'w'}$ in the sub-cloud layer with a much weaker magnitude. This local maximum $\overline{w'w'}$ occurs near the surface during daytime and halfway between the surface and cloud level during the nighttime. Although LR does not suffer from a particularly strong over-entrainment bias \cite{parishani2017toward}, the cloud layer is supported rather unrealistically by a weak $\overline{w'w'}$ maximum \cite{hignett1991observations,heinze2015second,mechem2012thermodynamic}. 

HR has a much smaller cloud fraction and LWP than the other simulations (Figure \ref{fig:htww}b vs. others), and several symptoms implicate too much entrainment as a key cause. For instance, HR has a much warmer sub-cloud layer temperature (Figure \ref{fig:Tdiff}) and this is systematic across Sc regions (Figures~\ref{fig:tdiff_WAustralia} and \ref{fig:tdiff_Namibian}). 
The warming cannot readily be explained by a difference in surface fluxes, given that HR's surface heat fluxes are roughly identical (Figure \ref{fig:Tdiff}) to the other HR configurations, especially during the first five days. We view HR's warm sub-cloud layer as a symptom of over-entrainment of warm overlying free-tropospheric air: As a result of enhanced turbulence through and above the cloud layer in HR, upward water transport is unable to sustain the cloud against entrainment-driven warming and drying. In summary, we suspect that strong $\overline{w'w'}$ near cloud top leads to over-entrainment, and that this is the main cause of the dim bias over Sc regions in the baseline HR simulation. 

We now analyze our attempts to reduce over-entrainment, beginning with applying hyperviscosity in isolation. We expect this to directly reduce $\overline{w'w'}$ associated with small eddies regardless of whether they were associated with moist processes. Indeed, above the cloud layer, adding the hyperviscosity term with $\tau = 30,15$ s helps to reduce the above-cloud magnitude of $\overline{w'w'}$ (Figure \ref{fig:htww}c,d) compared with HR (Figure \ref{fig:htww}b). Encouragingly, the low cloud fraction also increases throughout the simulation. But the additional low cloud is only recovered at night, which explains why hyperviscosity alone is not able to alleviate the shortwave dim bias. Reducing $\tau$ from 30 to 15 s (HRh15) helps to further reduce $\overline{w'w'}$ (Figure \ref{fig:htww}d), but has minimal effects on low cloud fraction beyond those of HRh; indeed this is why we use $\tau = 30 s$ as our default value for the hyperviscosity term. As pointed out earlier, despite being encouraging, the effects of hyperviscosity alone are not enough to fully address the over-entrainment problem that causes HR to be unable to sustain enough low cloud. 
One signature of turbulence driven by healthy amounts of cloud top radiative cooling is an elevated peak in $\overline{w'w'}$ away from the surface; note that this is too weak in HRh. 

We now examine the impact of additionally including cloud droplet sedimentation (Figure \ref{fig:htww}f). Removing cloud water away from the cloud top via sedimentation results in a \textit{larger} $\overline{w'w'}$ and a promising improvement in overall low cloud fraction. As suggested by \citeA{bretherton2007cloud}, this is driven by a reduction of the entrainment efficiency due to reduced liquid in the cloud-top entrainment zone. Reassuringly, the strongest $\overline{w'w'}$ is now found well above the surface and in the upper half of the boundary layer, consistent with cloud-top buoyancy production being the primary driver of convection, especially during the nighttime as observed. Thick clouds persist beyond the first week of the simulation as a result of the reduced consumption of cloud liquid by entrainment. Again, these effects only occur in conjunction with hyperviscosity. Incorporating sedimentation on its own results in a reduced cloud fraction and a weaker $\overline{w'w'}$ (Figure \ref{fig:htww}e). 

 \begin{table} 
 \caption{Median values from day 2 to day 15 for selected variables during daytime with nighttime values in parenthesis. This calculation discards $10\%$ of the upper and lower outlier points before estimating the median values.   }
 \centering
 \begin{tabular}{c c c c c}
 \specialrule{.1em}{.05em}{.05em} 
    & LR & HR & HRh & HRhs15\\
 \hline
   Cloud fraction ($\%$)                & 44 (73) & 27 (46) &  31 (63) & 49 (73)\\
   $LWP (g/m^2)$                       & 19 (43) & 7 (19)  &  10 (13) & 24 (55) \\
   $z_{i}$ (m)                          & 1301 (1216) & 1414 (1358) &  1387 (1298) & 1266 (1212) \\
   Cloud top (m)                        & 1212 (1157) & 1218 (1283) &  1215 (1224) & 1195 (1195)\\
   Cloud base (m)                       & 732 (661)   & 752 (705)   &  753 (685)   & 653 (618) \\
   $z_{LCL}$ (m)                        & 630(591)    & 649 (624)   &  639 (602)   & 560 (548) \\
   $\alpha _{q}$                       & 0.24 (0.22) & 0.22 (0.27)  &  0.22 (0.26) & 0.23 (0.24) \\
   $w_{e}$ (mm/s)                      & 3.6 (3.9) & 4.3 (3.8)  &  4.3 (3.8) & 4.0 (3.8) \\   
   $SWCRE (W/s^2$)                      & 59 (-) & 23 (-)  &  32 (-) & 71 (-) \\   
 \specialrule{.1em}{.05em}{.05em} 
 \label{tab:AllCaseAvgValue}
 \end{tabular}
 \begin{tablenotes}
      \small
      $\bullet$ Note: $LWP$=Total grid-box cloud liquid water path; $z_{LCL}$=Lifting condensation level; $z_i$=Inversion height; $\alpha _{\theta}$=decoupling parameters for potential temperature; $\alpha _{q}$=Decoupling parameters for potential water vapor; $w_{e}$=Entrainment rate (estimated as the subsidence rate at z=zi, assuming a steady state); $SWCRE$=Shortwave cloud forcing (values are all negative during the daytime).
 \end{tablenotes}
 \end{table}
 
Differences in inversion height are subtle to detect visually, but Table \ref{tab:AllCaseAvgValue} summarizes median properties from cloudy grid points (nonzero liquid water) in each simulation (averaged between days 2 to 15). The inversion height ($z_i$) shown in Table \ref{tab:AllCaseAvgValue} generally agrees with what was observed during the Ocean–Cloud–Atmosphere–Land Study Regional Experiment (VOCALS-REx) at Point Alpha in October and November 2008 off the coast of South America, which varied between 996 m to 1450 m  \cite{dodson2021turbulent}. We note that HRhs15 has the smallest difference between the cloud top height and the inversion height while HR has the largest, which we connect to inversion strength in the next section. 

Table \ref{tab:AllCaseAvgValue} also quantifies the decoupling that was discussed subjectively in Figure \ref{fig:htww}. We estimated the decoupling parameters for potential water vapor ($q$) as $\alpha _{q} = \frac{q _{cld} - q _{ml}}{q _{inv}-q _{ml}}$ \cite{park2004new}. Subscripts $cld$, $inv$, and $ml$ refer to mean values between the cloud base and top, between the surface and cloud base, and at the inversion height respectively. A decoupling parameter close to zero indicates a well-mixed boundary layer. Previous observations suggest that the boundary layer is decoupled when the parameter exceeds about 0.30 \cite{albrecht1995marine}. HR produces the highest values for $\alpha _{q}$ during the nighttime. There is much less contrast in daytime than nocturnal decoupling in HRhs15 compared with other configurations. A vertically decoupled thermodynamic structure produces cloud bases well above the LCL \cite{miller1998diurnal}; indeed the cloud base is 103 $m$  (81 $m$) above the LCL for HR during the daytime (nighttime), while 93 $m$ (70 $m$) for HRhs15. In the baseline HR, more decoupling can be viewed as a symptom of over-entrainment that is likely to cause less Sc during the daytime in that reduced moisture supply at cloud base cannot overcome the dry air entrainment from cloud top. HRhs15 corresponds to a lower entrainment rate ($w_{e}$) during the daytime than HR and HRh. HRhs15 also corresponds to the largest magnitude of shortwave cloud effects (SWCRE) due to a larger cloud fraction. 

\begin{figure} 
    \centering
    \includegraphics[scale=0.6]{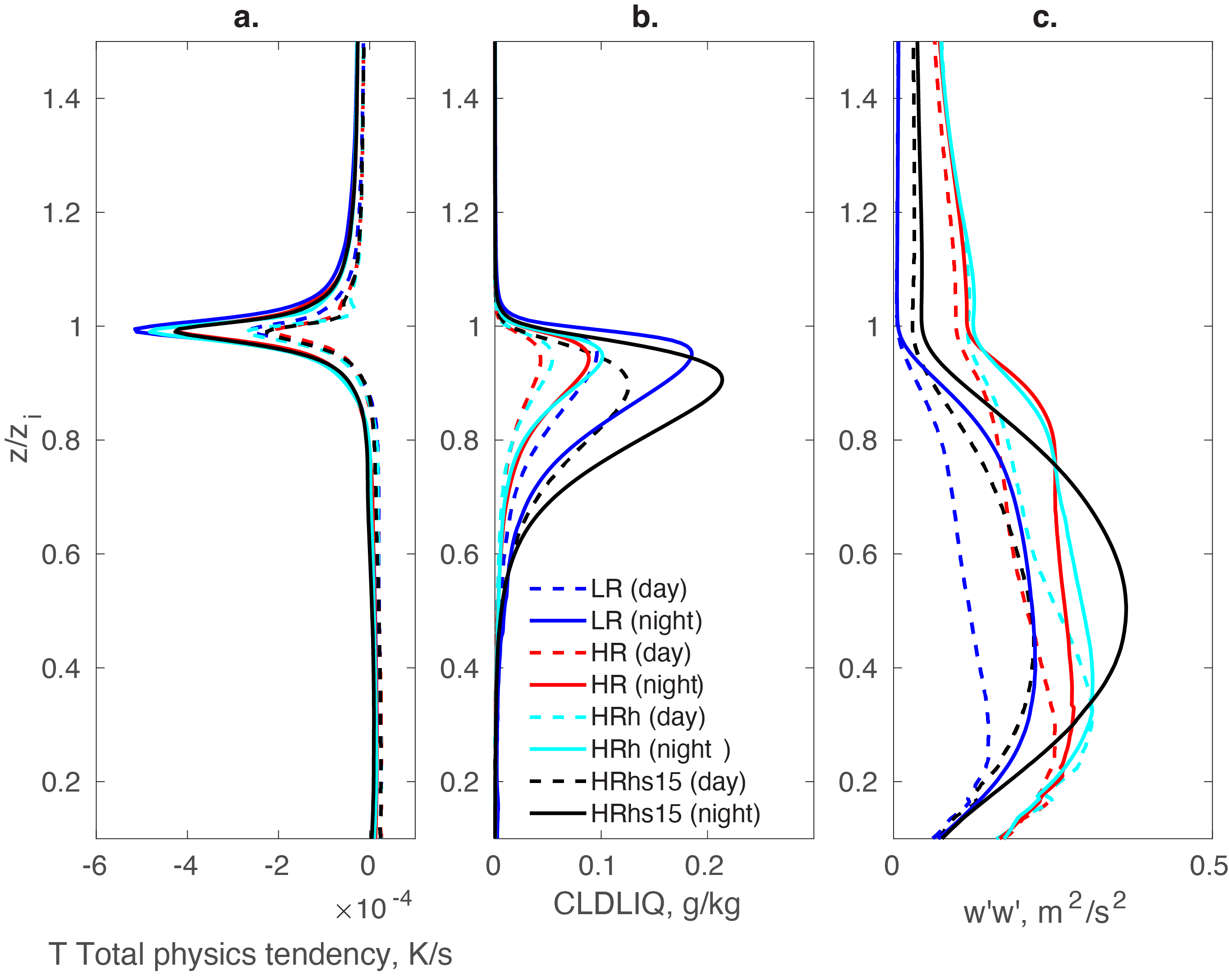}
    \centering
    \caption{Averaged vertical profiles nomalized by $z_{i}$ from day 2 to day 15 for (a) T total physics tendency ($K/s$), (b) cloud liquid water content (g/kg), and (c) $\overline{w'w'}$ ($m^2/s^2$). Solid lines represent night time average, while dashed lines represent the day time average. }
    \label{fig:Vprofile}
\end{figure}
 
 \subsection{Composite vertical structure and turbulent scale analysis}\label{Turbulence_scale}

We now examine mean daytime (dashed) and nighttime (solid) vertical profiles (Figure \ref{fig:Vprofile}) from days 2 to 15 over the Peruvian region. Here, the height, $z$, is normalized by the inversion height ($z_i$) to give a nondimensional vertical coordinate, $z/z_i$. 

All configurations show a large diurnal cycle: the daytime cloud liquid water content is about half of its nighttime value. HRhs15 (HR) corresponds to the largest (lowest) cloud liquid water content in both nighttime and daytime groups. It is interesting to note that HRhs15 has an even higher daytime cloud liquid water content than nighttime HR; in fact, in the following section we will show that there is too much daytime liquid in HRhs15, motivating compensatory microphysical retuning.

Longwave radiative cooling at the cloud top is regarded as the primary driver of convection in stratocumulus clouds \cite{lilly1968models,nicholls1989structure,moeng1996simulation}. Note that the peak cloud top radiative cooling that has the largest contribution to the total temperature tendency (Figure \ref{fig:Vprofile}a) occurs at the cloud top \cite{vanzanten2002radiative}. Regardless of the large magnitude of $\overline{w'w'}$ (Figure \ref{fig:Vprofile}c), HRhs15 has slightly lower cloud top radiative cooling compared with LR and HRh (Figure \ref{fig:Vprofile}a) due to less liquid emissivity near the cloud top. The level corresponding to the highest cloud liquid water content (Figure \ref{fig:Vprofile}b) is similar for LR, HR, and HRh, but this level is lower and further away from the cloud top for HRhs15 due to sedimentation. While HRhs15 and HR have a similar magnitude of cloud top radiative cooling, this radiative cooling produces a thicker cloud and drives stronger and better coupled vertical motions $\overline{w'w'}$ in HRhs15 than HR due to its weaker entrainment. A smaller magnitude of cloud top radiative cooling for HR during the daytime might also help to explain the warmer sub-cloud layer temperature compared with HRh, especially after day 7 shown in Figure \ref{fig:Tdiff}a. On the other hand, this warming is less severe compared between HR and HRhs15 near the cloud top.   

\begin{figure} 
    \centering
    \includegraphics[scale=0.50]{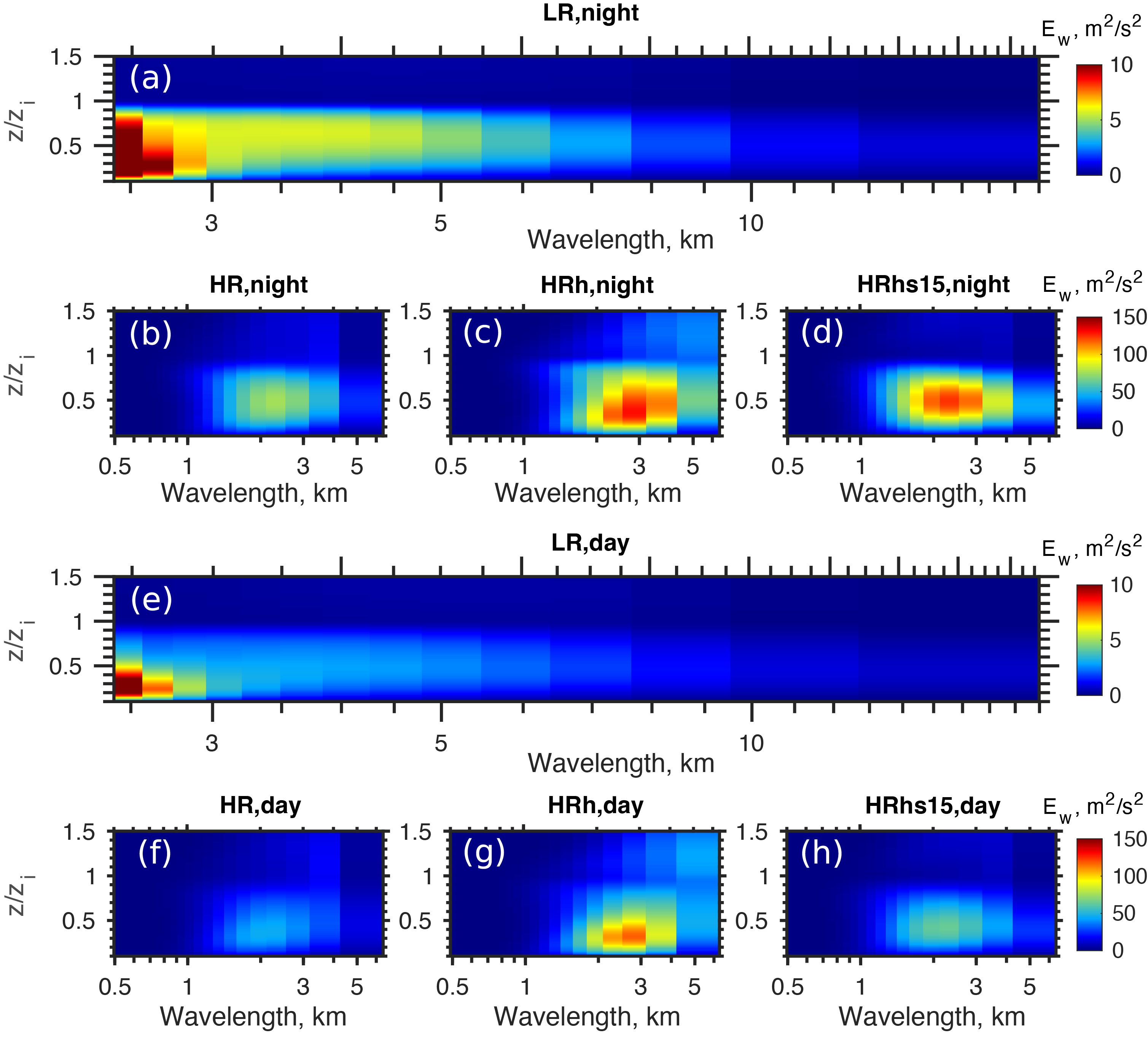}
    \centering
    \caption{A comparison of vertical profiles of the spectral intensity for nighttime average of (a) LR, (b) HR, (c) HRh, and (d) HRhs15, and daytime average of (e) LR, (f) HR, (g) HRh, and (h) HRhs15.  }
    \label{fig:SPowerDist}
\end{figure}

 It is still not clear whether cloud top entrainment is controlled by small eddies or large eddies \cite{wood2012stratocumulus}. To more closely examine the eddy spectra in cloudy regions, we perform spectral analysis along the CRM's horizontal dimension looking at the power spectrum of the CRM-scale vertical velocity, separately for daytime and nighttime (Figure \ref{fig:SPowerDist}). 

Compared with HR (Figure \ref{fig:SPowerDist}bf), the spectral intensity distribution for HRhs15 (rightmost column) is clearly confined within the stable boundary layer (STBL) with a single peak near 0.5$z_i$.The magnitude of this peak is larger during the nighttime since nighttime has a larger LWP. In HRh, the spectral intensity distribution peaks at towards larger wavelengths (Figure \ref{fig:SPowerDist}cg). In HRh, the spectral intensity of small  eddies (wavelengths close to $2\Delta x =  400$ $m$) is reduced, while the spectral power above the cloud top is enhanced with the large eddies (wavelengths greater than 2000 $m$). HRh has the strongest signals close to the largest wavelength that can be resolved (half of the domain size). LR is able to resolve eddies with horizontal wavelengths of up to ~10 km thanks to its larger domain size (Figure \ref{fig:SPowerDist}ae). However, the occurrence of the peak eddy spectral density for the smallest resolvable eddies (2.4 km) indicates eddy variance pile-up on the grid-scale. This implies that cloud-forming eddies, which are constrained by the numerical grid's finest resolved scale, are under-resolved in the low resolution simulation. 
The HR configurations are more physically plausible representations of sub-cloud turbulence in that a spectral peak exists interior to the resolvable scales. The high resolution of HR permits sub-cloud eddies to occupy multiple horizontal grid columns. We still can see relatively strong eddies (wavelengths around 3000 $m$) above the cloud. While HRhs15 may also cut off signals due to limited domain size, its spectral intensity distribution clearly shows a much better range of the resolved signal and STBL structure that is more consistent with observations and expectations from LES.  

HRhs15 also corresponds to the weakest above cloud eddies to reduce entrainment. Close to the inversion height ($z_i$), a stronger stratification tends to reduce the magnitude of $E_w$ and the dominant wavelength corresponding to the maximum $E_w$. Unlike HRhs15, relatively large vertical velocity fluctuations above the inversion height in HR and HRh corresponds well with a reduce the temperature inversion and, therefore, does not have a significant impact on the wavelength near the cloud top (e.g. Figure \ref{fig:SPowerDist}b). Above $z_i$, the eddies are much larger than they are below $z_i$.  

\begin{figure} 
    \centering
    \includegraphics[scale=0.42]{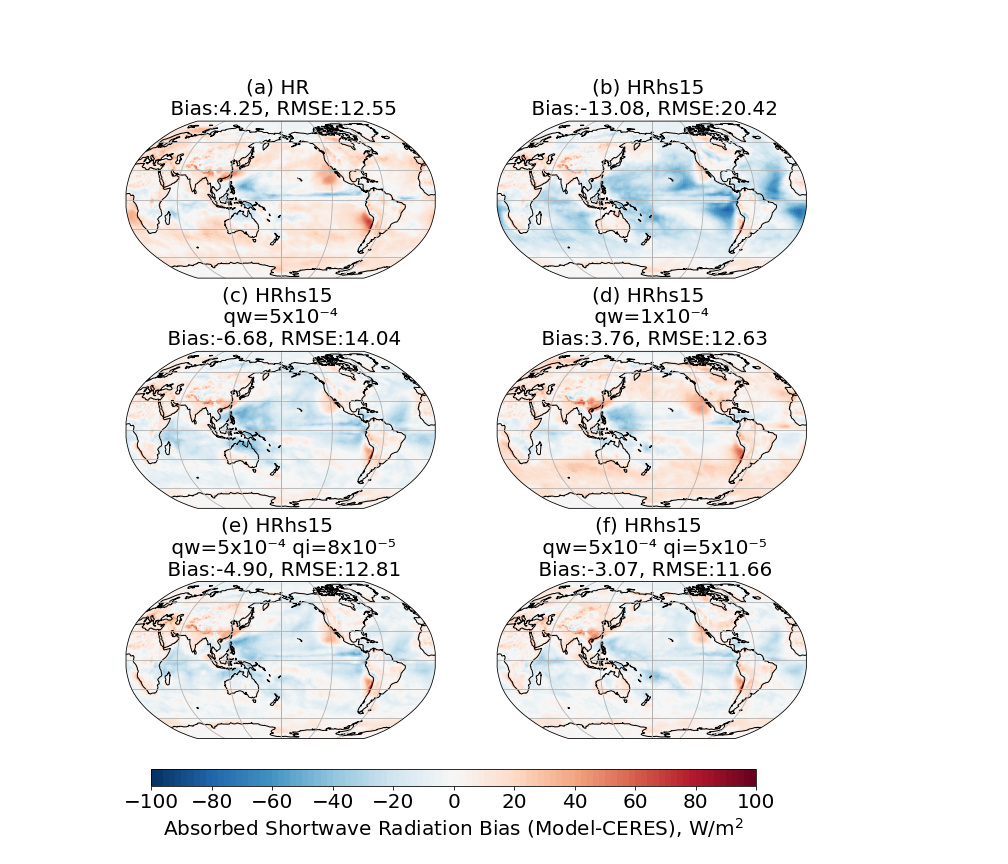}
    \centering
    \caption{Absorbed shortwave radiation at TOA biases with respect to CERES for the (a) HR, (b) HRhs15 with default configuration, (c) HRhs15 with $q_{w}=5\times 10^{-4}$,(d) HRhs15 with $q_{w}=1\times 10^{-4}$, (e) HRhs15 with $q_{w}=5\times 10^{-4}$ and $q_{i}=8\times 10^{-5}$ (d) HRhs15 with $q_{w}=5\times 10^{-4}$ and $q_{w}=5\times 10^{-5}$ }
    \label{fig:FSNTOAbias}
\end{figure}

 \subsection{Microphysics Tuning}\label{MicroTuning}

The previous sections have provided evidence that the addition of hyperviscosity and cloud droplet sedimentation produces encouraging changes in stratocumulus clouds such as more daytime clouds. However, the coarse global mesh, smaller CRM domains and short, two week duration of the hindcast simulations analyzed above represent notable compromises when compared to standard configurations of E3SM and E3SM-MMF. To address this concern we have conducted a series of 6-month simulations using the ne30pg2 grid and a larger CRM domain (Section \ref{Experiments}) that is more typical for E3SM experiments. These simulations are a subset of a larger tuning effort that considered several microphysical parameters. Ultimately, we found that autoconversion thresholds for liquid ($q_{cw0}=1\times 10^{-3}$ by default) and ice ($q_{ci0}=1\times 10^{-4}$ by default) were the most effective parameters for bringing the TOA energy fluxes into a reasonable balance and so that is what we will focus on below. This tuning exercise was partially motivated by the observation that the low-cloud enhancement resulting from the use of hyperviscosity and cloud droplet sedimentation produced a dramatic change in the TOA net shortwave radiative flux (Figure  \ref{fig:FSNTOAbias}b). 

 \begin{table} 
 \caption{The bias and RMSE in parenthesis for  ocean and land}
 \centering
 \begin{tabular}{c c c c c c c}
 \specialrule{.1em}{.05em}{.05em} 
  \multirow{2}{*}{ } &
  $q_{cw0}$&
  $q_{ci0}$ &
  \multicolumn{2}{c}{ASR (W m$^{-2}$)} &
  \multicolumn{2}{c}{OLR (W m$^{-2}$)} \\
 \cline{4-7}
    &  (kg kg$^{-1}$) &  (kg kg$^{-1}$) & ocean & land & ocean & land\\
 \hline
   HR  & 1$\times$10$^{-3}$ & 1$\times$10$^{-4}$
   &  4.0(9.5)  &  1.5(4.4) &  -8.2(8.6) & -2.0(2.8) \\
   HRhs15  & 10$^{-3}$ & 1$\times$10$^{-4}$ 
   &-19.5(18.6) & -2.1(4.3) &  -7.7(8.1) &  1.5(3.2) \\
   HRhs15  & 5$\times$10$^{-4}$ & 1$\times$10$^{-4}$                       &-10.4(11.3) & -1.1(4.4) &  -5.9(7.0) &  0.1(3.2) \\
   HRhs15  & 1$\times$10$^{-4}$ & 1$\times$10$^{-4}$ 
   &  3.0(9.7)  & 2.6(4.4)  &  -3.9(6.2) & -1.1(2.7) \\
   HRhs15  & 5$\times$10$^{-4}$ & 8$\times$10$^{-4}$ 
   & -8.6(10.3) & -0.2(4.1) &  -4.8(6.1) &  1.3(3.1) \\
   HRhs15  & 5$\times$10$^{-4}$ & 5$\times$10$^{-4}$ 
   & -6.9(8.9)  & 1.8(4.1)  &  -2.8(4.7) &  2.9(2.8) \\
 \specialrule{.1em}{.05em}{.05em} 
 \label{tab:SROLR_ocnlnd}
 \end{tabular}
 \end{table}

Figure \ref{fig:FSNTOAbias} shows a comparison of the absorbed shortwave radiation climatological biases from our ten-year simulation compared with satellite observations. The baseline HR and several retunings of the baseline HRhs15 configuration are compared. Unlike HR, the ASR bias for HRhs15 primarily stems from too bright marine clouds, especially over the subtropical Sc regions (Figure \ref{fig:FSNTOAbias}). Especially strong negative ASR biases are found off the western coasts of Peru, Namibia, Australia, and California. Reducing the liquid autoconversion thresholds increases the ASR (Figure \ref{fig:FSNTOAbias} bcd), approximately halving the global mean shortwave bias, and reduces the RMSE. A reduced liquid autoconversion threshold combined with an increased ice autoconversion threshold further helps to ameliorate ASR global mean bias and reduce RMSE (Figure \ref{fig:FSNTOAbias}ef). The configuration with $q_{cw0}=5\times 10^{-4}$ and $q_{ci0}=5\times 10^{-5}$ produces the smallest global mean bias and RMSE (Figure \ref{fig:FSNTOAbias}e). Most of this improvement occurs over the ocean (Table \ref{tab:SROLR_ocnlnd}). Even with the microphysics retuning, the incorporation of hyperviscosity and sedimentation continues to play a crucial role in reducing the ASR bias particularly over regions covered by stratocumulus clouds (Figure \ref{fig:Figure8mt}). 

\begin{figure} 
    \centering
    \includegraphics[scale=0.42]{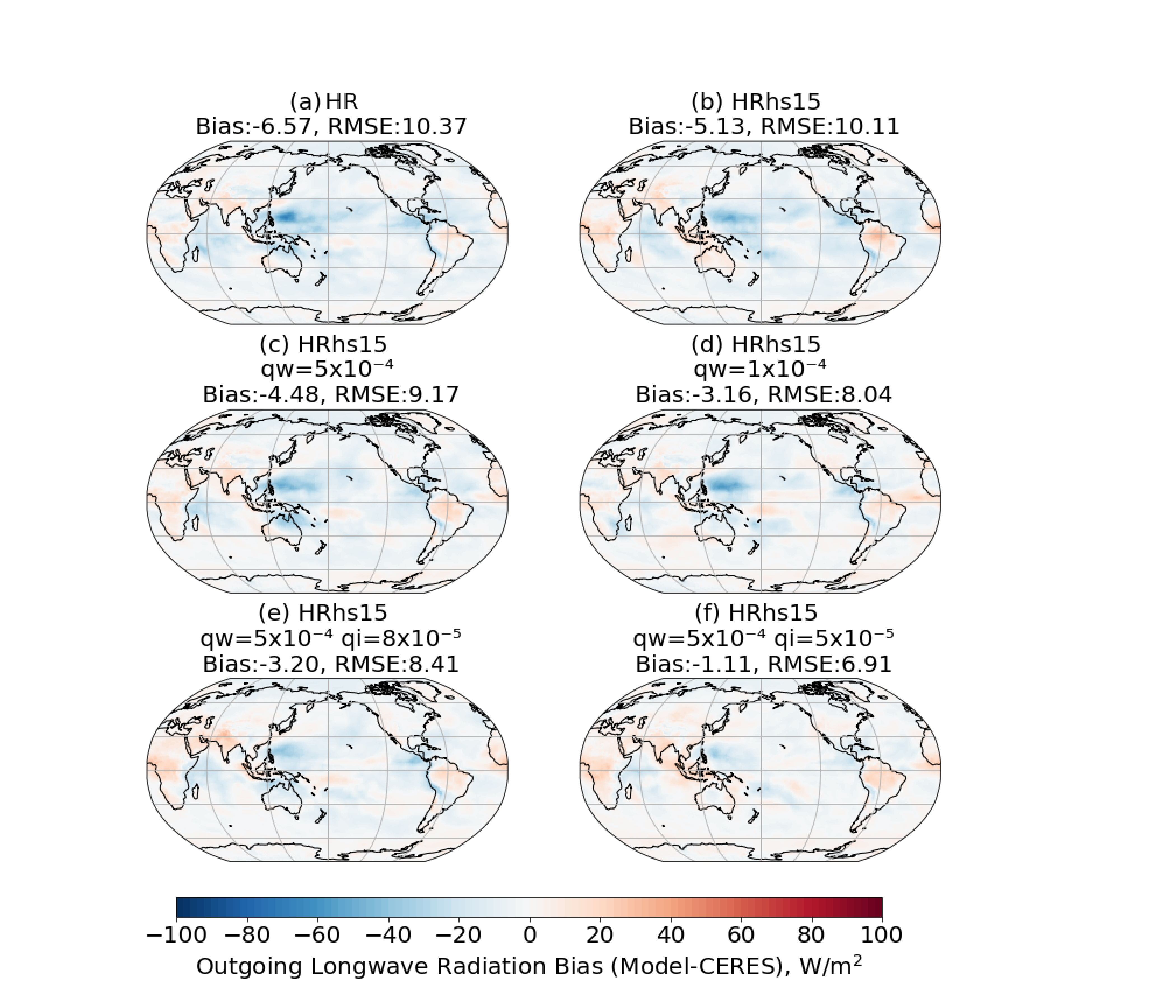}
    \centering
    \caption{Outgoing longwave radiation at TOA biases with respect to CERES for the (a) HR, (b) HRhs15 with default configuration, (c) HRhs15 with $q_{w}=5\times 10^{-4}$,(d) HRhs15 with $q_{w}=1\times 10^{-4}$, (e) HRhs15 with $q_{w}=5\times 10^{-4}$ and $q_{i}=8\times 10^{-5}$ (d) HRhs15 with $q_{w}=5\times 10^{-4}$ and $q_{w}=5\times 10^{-5}$}
    \label{fig:FLNTbias}
\end{figure}

Microphysical tuning results in weaker changes of OLR than those for ASR (Figure \ref{fig:FLNTbias}). Sedimentation (HRhs15) only slightly increased the global mean OLR (Figure \ref{fig:FLNTbias}ab) from the base HR simulation, which was too opaque in the tropics. This bias is reduced in Figure \ref{fig:FLNTbias}f. Overall, we are able to obtain less than $1 W/m^{2}$ OLR bias. 

 \section{Discussion and Conclusions}\label{DiscussionConclusions}

Compared to other available global modeling tools for studying cloud feedback, today's GPU-accelerated Multiscale Modeling Frameworks configured with High Resolution (HR) interior grids have the capacity to provide a unique combination of global eddy-permitting resolution coverage and multi-decadal throughput that complements other climate simulation technology. In theory HR MMFs should be attractive for low cloud feedback analysis and cloud-aerosol interactions, by making minimal assumptions about the sub-km scale vertical eddy field.

But in practice, this depends on the model's ability to represent present-day climate. For over five years since the first experiments with HR MMF, it has been unclear whether a chronic over-entrainment bias preventing realistic amounts of Sc liquid water was surmountable. It has been natural to wonder if the inherent idealizations of MMFs that make them computationally attractive -- i.e. the limited domain size, dimensionality, moderate (200-m) interior horizontal resolution, lateral periodicity, and associated inability to laterally advect liquid water conservatively -- \cite{muller2012detailed,jansson2022representing}  -- might impose fundamental limitations.

Our results argue otherwise: We suggest MMFs are simply in their infancy and their interior resolved scale has never been sufficiently tuned to succeed in a HR limit. To show this, we investigated the impact of adding hyperviscosity and sedimentation on low cloud formation in a high resolution multi-scale modeling framework (HR) that uses 200-m horizontal, and as fine as 20-m vertical, grid spacing within each of its embedded convection resolving models, configured with bulk one-moment microphysics. As in previous studies, our control HR simulation produced the familiar bias of too few low clouds over regions of subtropical marine stratocumulus (Sc), resulting in a dim bias compared with satellite observations of shortwave radiation absorbed at top of atmosphere. 

We found promising Sc-selective brightening when we combined scale selective damping (hyperviscosity) of grid scale momentum variations with the introduction of cloud droplet sedimentation. The application of hyperviscosity alone, which directly reduces $\overline{w'w'}$, leads to short-lived increases in nocturnal cloud thickness. Simulations with sedimentation alone lead to modest increases in cloud fraction.  However, the most encouraging effects occur when the two are applied together, whereupon robust increases in cloud liquid water lead to a reversal of the Sc dim bias, including in multi-week integrations. These nonlinear interactions of these two processes lead to much stronger changes than when they are applied separately.  The resulting, larger peak liquid water concentration is shifted downward, away from the cloud-top entrainment zone. In this configuration, dense, locally-formed Sc are observed to form in the HR MMF, and the sub-cloud eddy spectrum becomes especially well organized. 

In summary, with only these minor, physically-motivated re-tunings, the CRMs of a HR MMF can be coerced to create healthy amounts of locally generated stratocumulus liquid, in association with reasonable sub-cloud eddy properties. This is possible despite the assumptions of periodicity, dimensionality (2D) and only 200-m horizontal resolution that makes HRs computationally efficient, which is encouraging.  At first, the interventions create too much low cloud, and swap a regional dim bias for a global ocean bright bias -- but with encouragingly little horizontal variation across the oceanic cloud regimes, with hope for calibration. Thus, as must occur following any manipulation of a MMF's physical formulation, we performed a compensatory microphysical re-tuning to recover a reasonable top of atmosphere climatology. Despite a limited tuning campaign, the results demonstrate the potential for significantly less severe Sc dim biases, and reduced spatial RMSE of shortwave absorbed radiation across the global ocean. It is logical to expect that with further attention to tuning, even more operationally attractive configurations could be uncovered. 

Several limitations of this work are worth mentioning. In our current study, we did not fully explore the contribution of resolved scale advection and subgrid scale diffusion to the reduced entrainment efficiency caused by hyperviscosity. With enhanced outputs including the SGS contribution in SAM, future investigations could further evaluate these aspects to gain a deeper understanding of the underlying mechanisms. We speculate a root problem motivating the need to apply hyperviscosity in our simulations may be the numerics of the momentum solver in the embedded CRM (See Section \ref{ModelDes}). Successors to SAM under development by DOE for use in E3SM-MMF, like most modern LES \cite{eldred2021structure}, intentionally use a cell-centered, entropy stable Weighted Essentially Non-Oscillatory (WENO) schemes for the momentum solver, as suggested in the work of \cite{pressel2017numerics}. Another obvious limitation is that the HR here uses a simple one-moment microphysics scheme and diagnostic turbulence scheme. While this is helpful for maximizing throughput at its ambitious grid resolution, it is also outdated. The eventual higher-order microphysics \cite{morrison2015parameterization} that are expected to come online in the E3SM-MMF may suffer less from baseline over-entrainment due to already including a representation of the sedimentation process that we have argued helpfully draws cloud liquid down from the inversion to optimize entrainment efficiency. Perhaps this imminent next generation of HR will have less need for compensatory CRM-scale tuning and less sensitivity to grid spacing to achieve its low cloud potential. In our current study, we did not fully explore the contribution of resolved scale advection and subgrid scale diffusion to the reduced entrainment efficiency caused by hyperviscosity. With the enhanced outputs of the SGS contribution in SAM, future investigations could further evaluate these aspects to gain a deeper understanding of the underlying mechanisms.   
Several limitations of this work are worth mentioning. In our current study, we did not fully explore the contribution of resolved scale advection and subgrid scale diffusion to the reduced entrainment efficiency caused by hyperviscosity. With enhanced outputs including the SGS contribution in SAM, future investigations could further evaluate these aspects to gain a deeper understanding of the underlying mechanisms. We speculate a root problem motivating the need to apply hyperviscosity in our simulations may be the numerics of the momentum solver in the embedded CRM (See Section \ref{ModelDes}). Successors to SAM under development by DOE for use in E3SM-MMF, like most modern LES \cite{eldred2021structure}, intentionally use a cell-centered, entropy stable Weighted Essentially Non-Oscillatory (WENO) schemes for the momentum solver, as suggested in the work of \cite{pressel2017numerics}. Another obvious limitation is that the HR here uses a simple one-moment microphysics scheme and diagnostic turbulence scheme. While this is helpful for maximizing throughput at its ambitious grid resolution, it is also outdated. The eventual higher-order microphysics \cite{morrison2015parameterization} that are expected to come online in the E3SM-MMF may suffer less from baseline over-entrainment due to already including a representation of the sedimentation process that we have argued helpfully draws cloud liquid down from the inversion to optimize entrainment efficiency. Perhaps this imminent next generation of HR will have less need for compensatory CRM-scale tuning and less sensitivity to grid spacing to achieve its low cloud potential. In our current study, we did not fully explore the contribution of resolved scale advection and subgrid scale diffusion to the reduced entrainment efficiency caused by hyperviscosity. With the enhanced outputs of the SGS contribution in SAM, future investigations could further evaluate these aspects to gain a deeper understanding of the underlying mechanisms.   

Then again, perhaps not. For now, it is clear that ``multi-scale'' modeling frameworks seem to merit careful ``multi-scale'' physics calibration, and that this has largely been overlooked on the interior resolved scale, at least in explorations of MMF at the limit of HRs' grid resolutions. On the one hand, this annoyingly complicates the art of global model tuning. On the other hand, it is good news for the long term potential to study low-cloud feedbacks quasi-explicitly via the HR MMF approach. Despite its idealizations, healthy amounts of present-day Sc cloud can evidently be recovered in an HR MMF, allowing its computational advantages to be brought to bear on questions of cloud feedback. It will be important to determine whether this modifies previous estimates of the HR MMF low cloud feedback to warming \cite{parishani2018insensitivity} from previous generation simulations that have struggled to capture sufficient baseline low cloud.

\section*{Open Research}

All E3SM source code can be accessed from the GitHub repository of the \citeA{Project_Energy_Exascale_Earth}. All code modifications needed to implement our approach within a legacy fork of the E3SM MMF climate model can be found in the repository of \citeA{Peng_E3SM_with_hyperviscosity}. The raw output data is archived and can be accessed from Zenodo under the references of \cite{peng2023part1} and \cite{peng2023part2}.

\acknowledgments
L. Peng, M. Pritchard, P. N. Blossey, and C. S. Bretherton acknowledge funding from the National Science Foundation (NSF) Climate and Large-scale Dynamics program under grants AGS-1912134 and AGS-1912130. M. Pritchard, L. Peng, and A. M. Jenney further acknowledge DOE BER (DE-SC0023368). The authors acknowledge the Texas Advanced Computing Center (TACC) at The University of Texas at Austin for providing {HPC, visualization, database, or grid} resources that have contributed to the research results reported within this paper. URL: http://www.tacc.utexas.edu
This research was supported by the Exascale Computing Project (17-SC-20-SC), a collaborative effort of the U.S. Department of Energy Office of Science and the National Nuclear Security Administration and by the Energy Exascale Earth System Model (E3SM) project, funded by the U.S. Department of Energy, Office of Science, Office of Biological and Environmental Research.
This work was performed under the auspices of the U.S. Department of Energy by Lawrence Livermore National Laboratory under Contract DE-AC52-07NA27344.
This research used resources of the Oak Ridge Leadership Computing Facility, which is a DOE Office of Science User Facility supported under Contract DE-AC05-00OR22725.

\appendix
\section{Figures}\label{AFig}

\begin{figure} 
    \centering
    \includegraphics[scale=0.5]{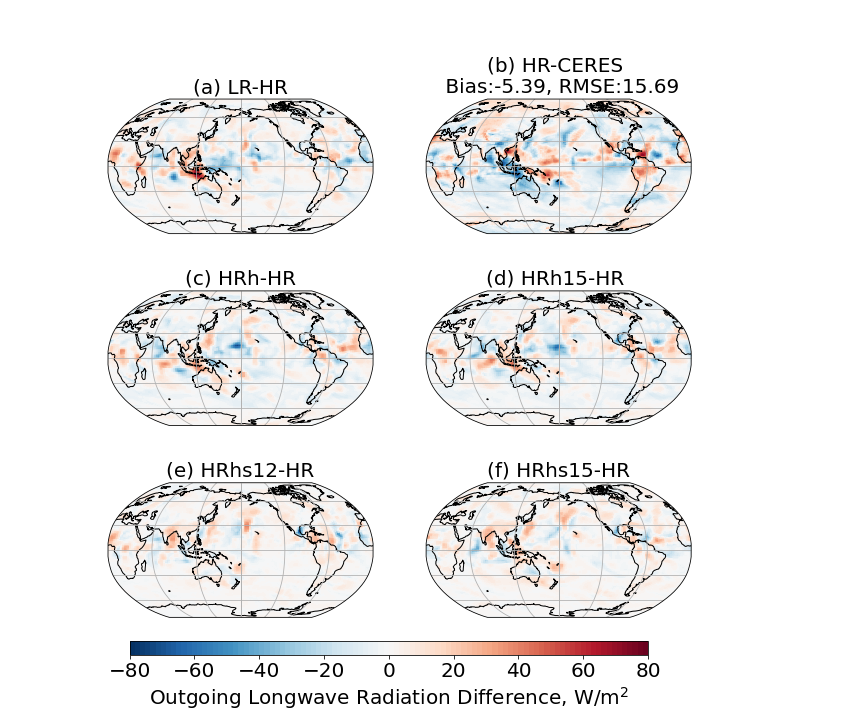}
    \centering
    \caption{Similar to Figure \ref{fig:asr} but for outgoing longwave radiation (OLR) difference. }
    \label{fig:olr}
\end{figure} 


\begin{figure} 
    \centering
    \includegraphics[scale=0.3]{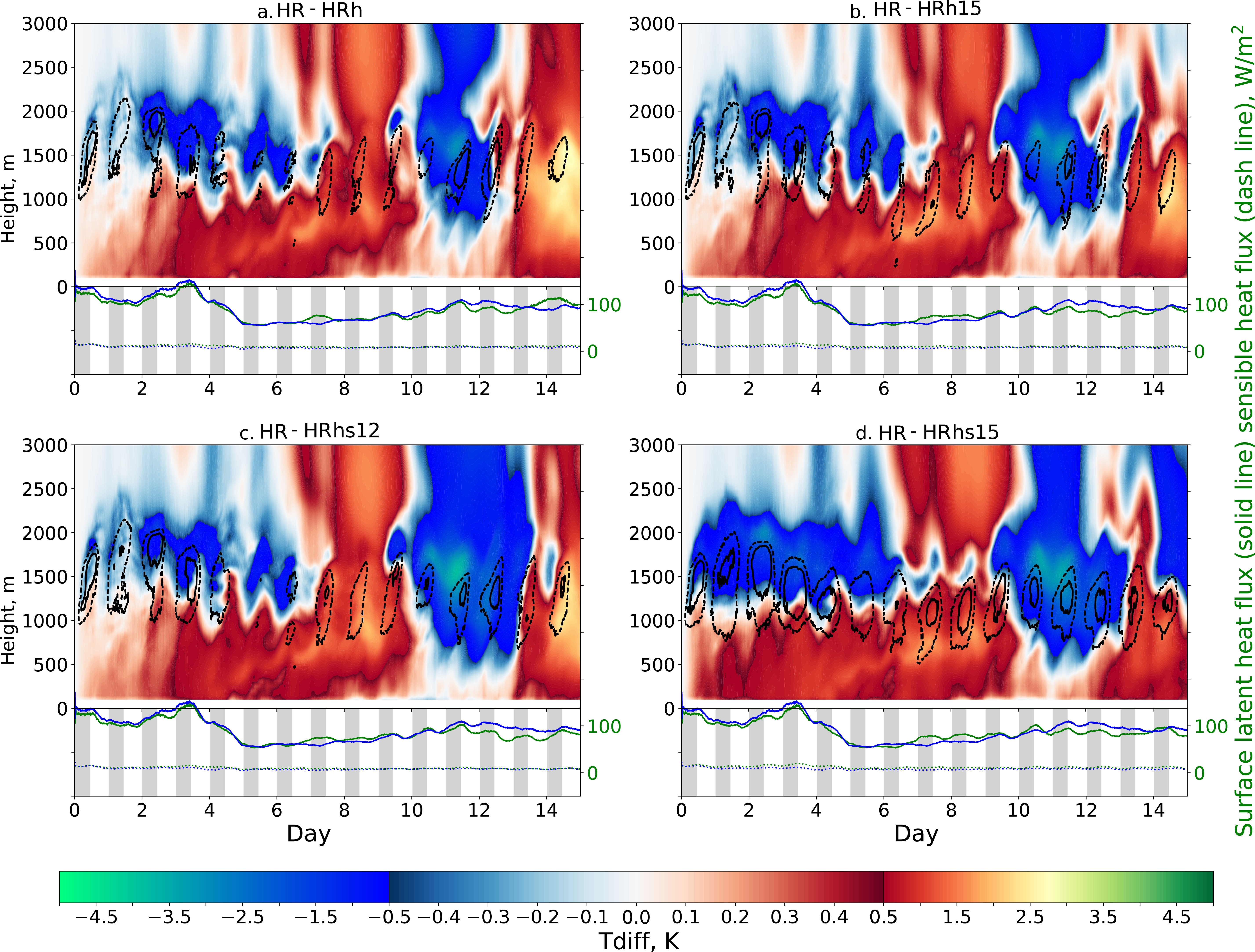}
    \centering
    \caption{Height time evolution of temperature difference over Peruvian between HR and (a) HRh, and (b) HRh15, (c) HRhs12, and (d) HRhs15. Surface sensible (dashed lines) and latent heat (solid lines) flux are shown by blue lines for HR and green lines for other configurations. Two cloud fraction contour lines for other configurations been subtracted by HR are shown for 0.1 (black dotted line) and 0.2 (black solid line).}
    \label{fig:Tdiff}
\end{figure}

\begin{figure} 
    \centering
    \includegraphics[scale=0.3]{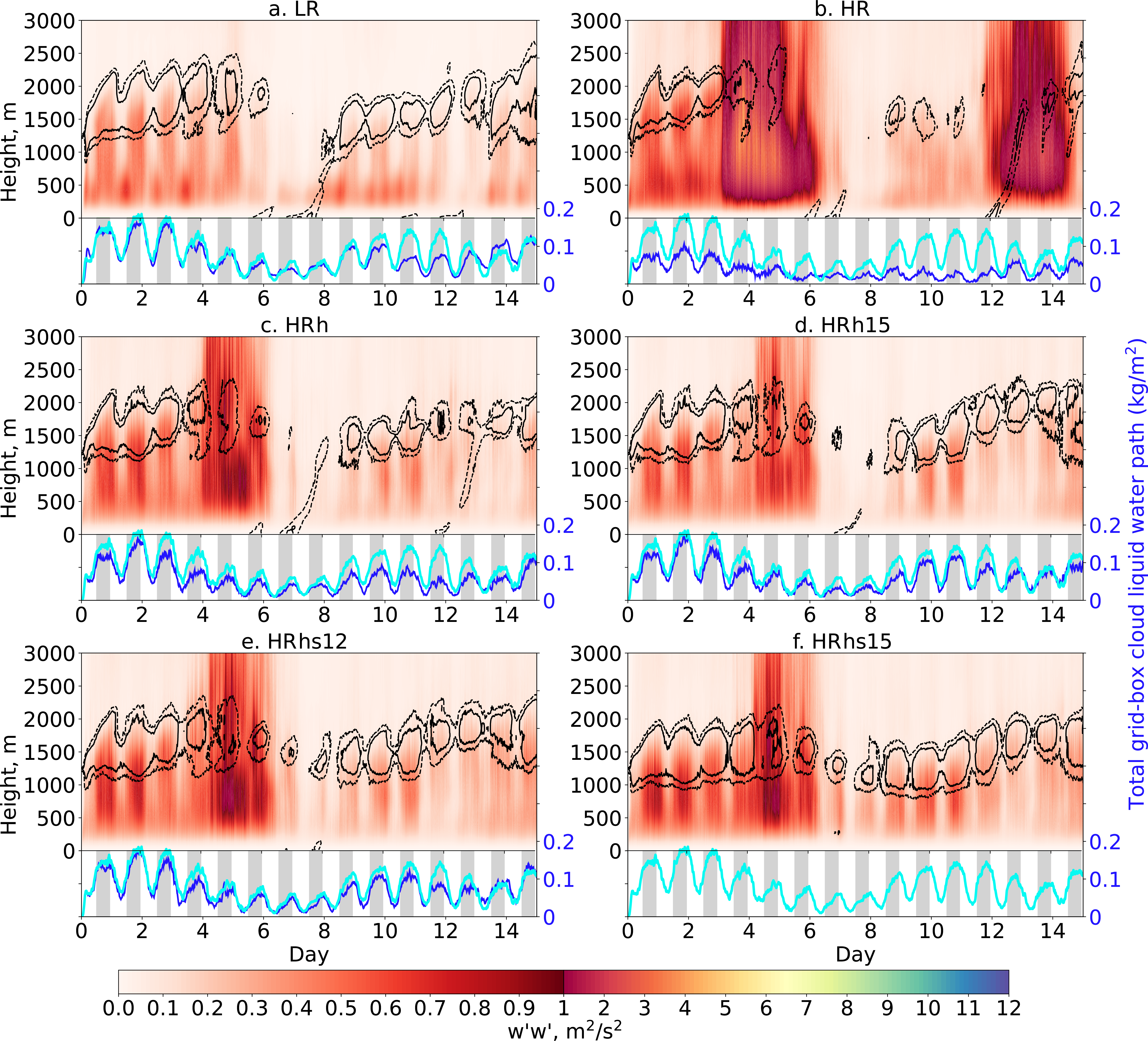}
    \centering
    \caption{Height time evolution of vertical velocity variance ($\overline{w'w'}$, in units of $m^2/s^2$) in West Australia averaged over 15 days starting from October 1st 2008. (a) LR, (b) HR, (c) HRh, (d) HRh15, (d) HRhs12, (f) HRhs15. The blue lines represents the total grid-box liquid water path and thick cyan line represents HRhs15 for all panels as a reference. Gray shaded time intervals represent nighttime. Black contours are showing the $10\%$ (black dashed line) and $20\%$ (black solid line) cloud fraction. }
    \label{fig:htww_WAus}
\end{figure}

\begin{figure} 
    \centering
    \includegraphics[scale=0.3]{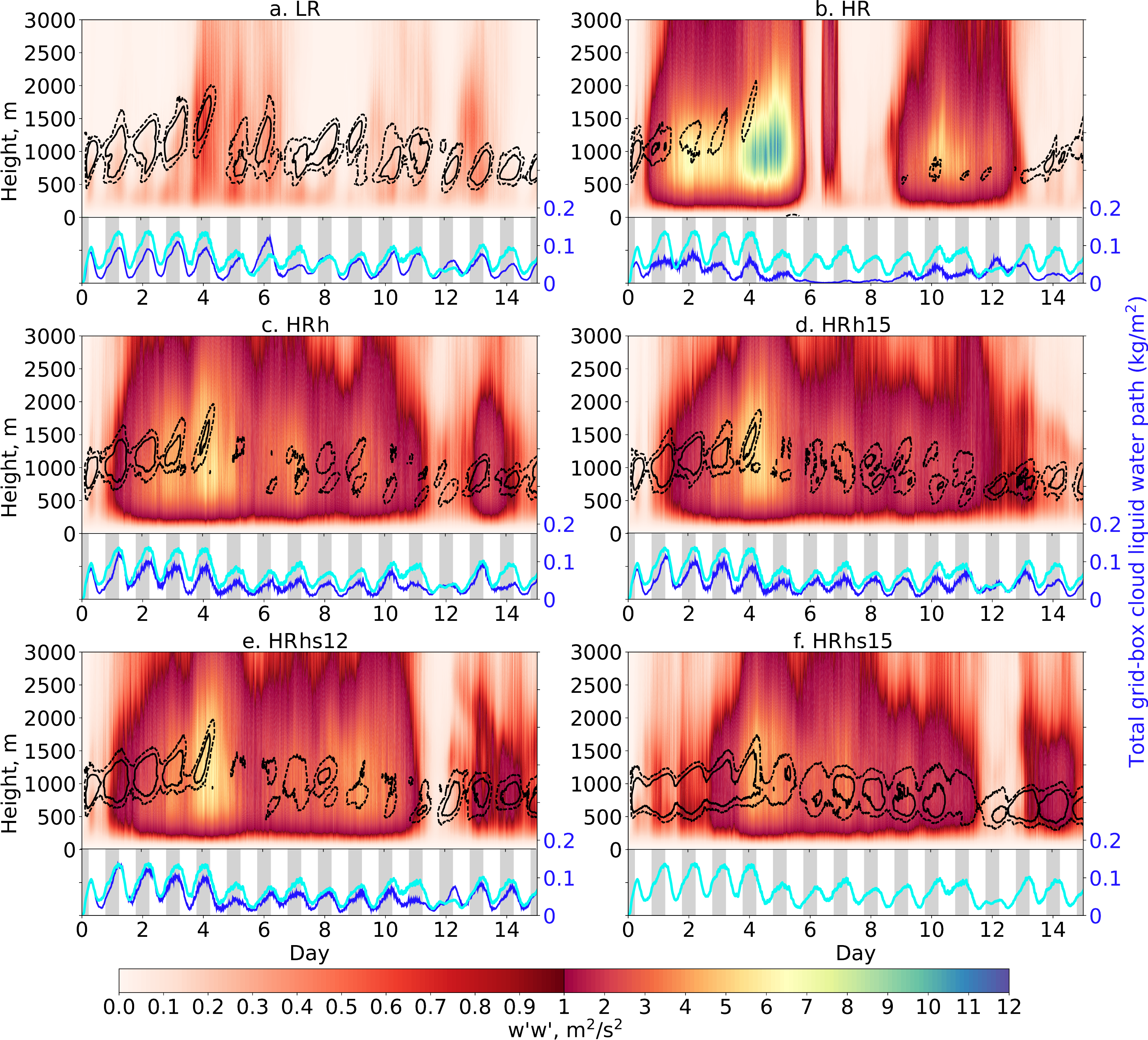}
    \centering
    \caption{Height time evolution of vertical velocity variance ($\overline{w'w'}$, in units of $m^2/s^2$) in Namibian averaged over 15 days starting from October 1st 2008. (a) LR, (b) HR, (c) HRh, (d) HRh15, (d) HRhs12, (f) HRhs15. The blue lines represents the total grid-box liquid water path and thick cyan line represents HRhs15 for all panels as a reference. Gray shaded time intervals represent nighttime. Black contours are showing the $10\%$ (black dashed line) and $20\%$ (black solid line) cloud fraction.}
    \label{fig:htww_Namibian}
\end{figure}

\begin{figure} 
    \centering
    \includegraphics[scale=0.3]{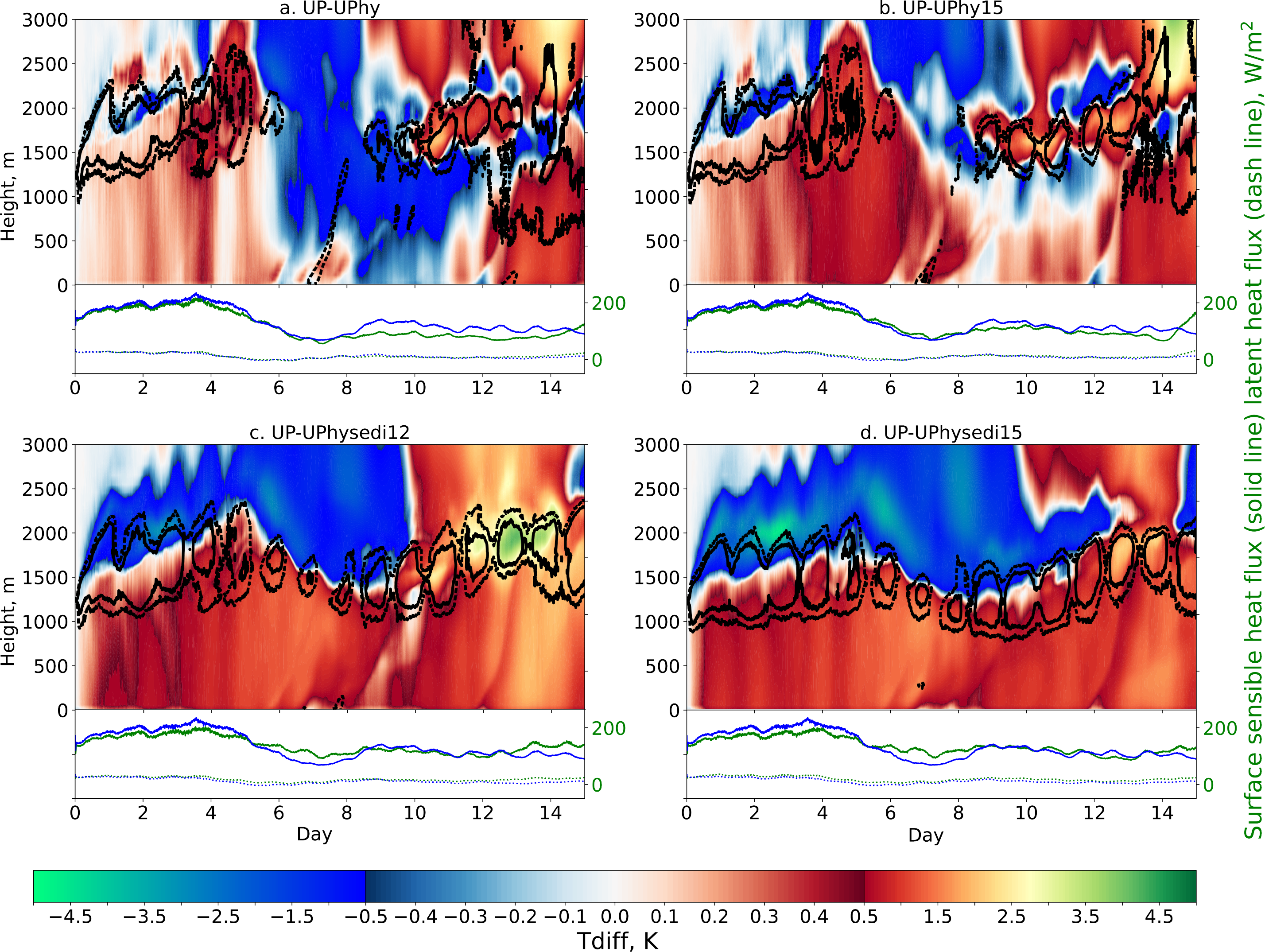}
    \centering
    \caption{Height time evolution of temperature difference in west coast Australia between HR and (a) HRh, and (b) HRh15, (c) HRhs12, and (d) HRhs15. Surface sensible and latent heat flux are shown by blue solid and dashed lines for HR (blue) and other configurations (green). Two cloud fraction contour lines are 0.1 (black dotted line) and 0.2 (black solid line).}
    \label{fig:tdiff_WAustralia}
\end{figure}

\begin{figure} 
    \centering
    \includegraphics[scale=0.3]{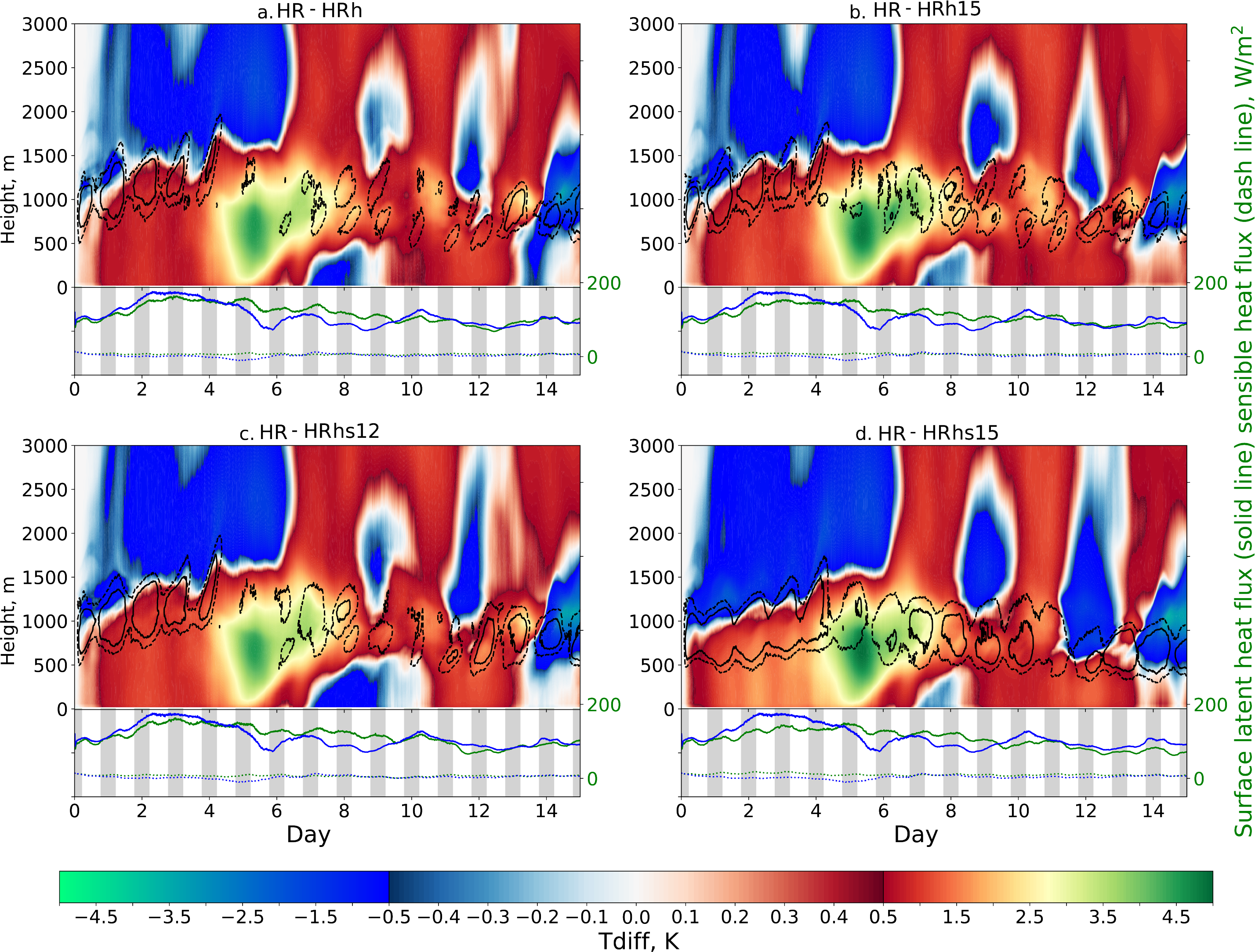}
    \centering
    \caption{Height time evolution of temperature difference in Namibian between HR and (a) HRh, and (b) HRh15, (c) HRhs12, and (d) HRhs15. Surface sensible and latent heat flux are shown by blue solid and dashed lines for HR (blue) and other configurations (green). Two cloud fraction contour lines are 0.1 (black dotted line) and 0.2 (black solid line).}
    \label{fig:tdiff_Namibian}
\end{figure}

\begin{figure} 
    \centering
    \includegraphics[scale=0.5]{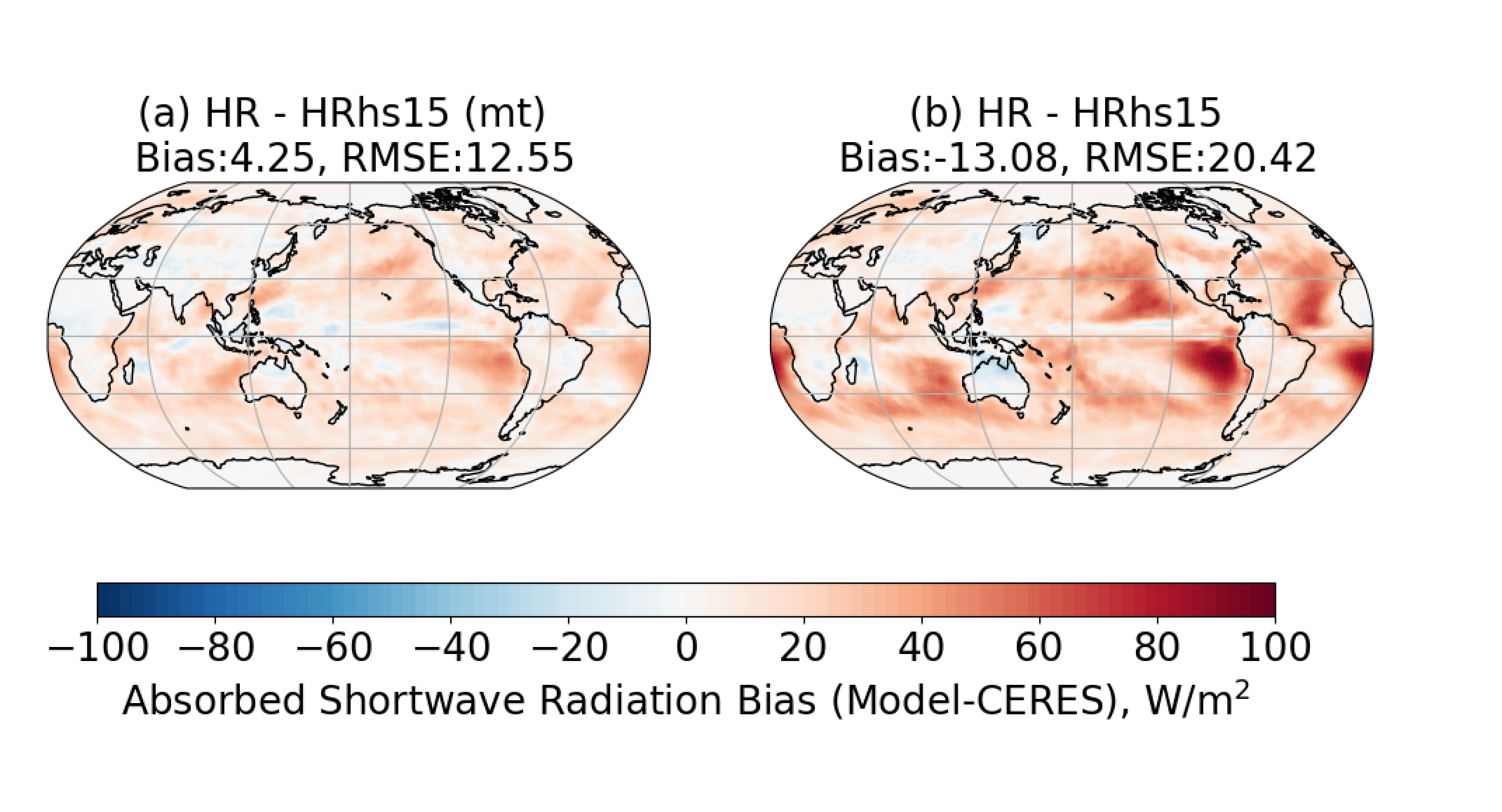}
    \centering
    \caption{The ASR bias differences between HR and HRhs15  (a) with microphysics tunning and (b) without microphysics tunning. }
    \label{fig:Figure8mt}
\end{figure}

\clearpage
\bibliography{ references }

%
%
%
%
%

\end{document}